\documentclass[a4paper,fleqn,usenatbib]{mnras}
\usepackage{mathtools}

\usepackage{breqn}
\usepackage[T1]{fontenc}
\usepackage{ae,aecompl}
\usepackage{tablefootnote}

\usepackage{graphicx}	
\usepackage{amsmath}	
\usepackage{amssymb}	
\usepackage{subfig}
\usepackage{float}
\usepackage{multirow}
\usepackage{units}
\usepackage{placeins}

\usepackage{xcolor}
\usepackage{ragged2e}

\usepackage{pdflscape}

\title[Discovery of LFQPO in Swift J1727.8-1613 in hard X-rays]
{Discovery of evolving low-frequency QPOs in hard X-rays ($\sim 100$ keV) observed in black hole Swift J1727.8-1613 with {\it AstroSat}}

\author[Nandi et al.]{Anuj Nandi$^{1}$, Santabrata Das$^{2}$\thanks{E-mail: sbdas@iitg.ac.in}, Seshadri Majumder$^{2}$, Tilak Katoch$^{3}$, H. M. Antia$^{4}$, Parag Shah$^{3}$\\
	$^{1}$Space Astronomy Group, ISITE Campus, U. R. Rao Satellite Centre, Outer
	Ring Road, Marathahalli, Bangalore, 560037, India.\\
	$^{2}$Department of Physics, Indian Institute of Technology Guwahati, Guwahati, 781039, India.\\
	$^{3}$DAA, Tata Institute of Fundamental Research, Colaba, Mumbai, 400005, India. \\
	$^{4}$UM-DAE Centre for Excellence in Basic Sciences, University of Mumbai, Kalina, Mumbai, 400098, India.
}

\date{Accepted XXX. Received YYY; in original form ZZZ}

\pubyear{2023}

\begin{document}
	\label{firstpage}
	\pagerange{\pageref{firstpage}--\pageref{lastpage}}
	\maketitle
		
	\begin{abstract}
		We report the first detection of evolving Low-Frequency Quasi-periodic Oscillation (LFQPO) frequencies in hard X-rays upto $100$ keV with {\it AstroSat/LAXPC} during `unusual' outburst phase of Swift J1727.8-1613 in hard-intermediate state (HIMS). The observed LFQPO  in $20 - 100$ keV has a centroid $\nu_{_{\rm QPO}}=1.43$ Hz, a coherence factor $Q= 7.14$ and an amplitude ${\rm rms_{_{\rm QPO}}} = 10.95\%$ with significance $\sigma = 5.46$. Type-C QPOs ($1.09-2.6$ Hz) are found to evolve monotonically during HIMS of the outburst with clear detection in hard X-rays ($80 - 100$ keV), where ${\rm rms_{_{\rm QPO}}}$ decreases ($\sim 12-3\%$) with energy. Further, $\nu_{_{\rm QPO}}$ is seen to correlate (anti-correlate) with low (high) energy flux in $2-20$ keV ($15-50$ keV). Wide-band ($0.7 - 40$ keV) energy spectrum of {\it NICER/XTI} and {\it AstroSat/LAXPC} is satisfactorily described by the `dominant' thermal Comptonization contribution ($\sim 88$ \%) in presence of a `weak' signature of disk emissions ($kT_{\rm in} \sim 0.36$ keV) indicating the harder spectral distribution. Considering source mass $M_{\rm BH}=10M_\odot$ and distance $1.5 < {\rm d~(kpc)} < 5$, the unabsorbed bolometric luminosity is estimated as $\sim 0.03-0.92\%L_{\rm Edd}$. Finally, we discuss the implications of our findings in the context of accretion dynamics around black hole X-ray binaries.
	\end{abstract}
	

\begin{keywords}
	accretion, accretion disc -- black hole physics -- radiation mechanism; general -- X-rays: binaries -- stars: individual: Swift J1727.8-1613
\end{keywords}

	
\section{Introduction}

Black hole X-ray binaries (BH-XRBs) are the ideal cosmic laboratories to understand the underlying physical mechanisms that govern the accretion dynamics around the compact objects. Interestingly, the accreting systems become more complex when BH-XRBs undergo various spectral state transitions during the outburst phases \cite[and references therein]{Belloni-etal2005,Remillard-McClintock2006,Nandi-etal2012,Iyer-etal2015,Sreehari-etal2019,Baby-etal2021,Prabhakar-etal2023}. Indeed, these sources often exhibit Low-Frequency QPO (LFQPO) features ($0.1$ to $\lesssim 30$ Hz) in the power spectra \cite[]{Remillard-McClintock2006}, which generally evolve during the onset phase of the outburst. In general, LFQPOs are considered as an effective diagnostic tool to examine the accretion scenarios \cite[]{Chakrabarti-Manickam2000,Chakrabarti-etal2008,Nandi-etal2012,Iyer-etal2015} and the nature of the central source \cite[]{Motta-etal2014a,Motta-etal2014b}. Usually, LFQPO features are observed at low energies ($\lesssim 40$ keV) \cite[and references therein]{Sreehari-etal2019,Aneesha-etal2023} except for very few sources ($i.e.$, GRS 1915+105: \cite{Tomsick-etal2001}, MAXI J1535$-$571: \cite{Huang-etal2018}, MAXI J1820$+$070: \cite{Ma-etal2021}, MAXI J$1803-298$: \cite{Wang-etal2021}, MAXI J$1631-479$: \cite{Bu-etal2021}), where LFQPO are observed at higher energies. Needless to mention that the detection of LFQPO signatures at higher energies is a rare phenomenon and its origin remains unclear till date.

Keeping this in mind, we examine the LFQPO features in hard X-rays observed with {\it AstroSat} for a X-ray transient source Swift J$1727.8-1613$ recently discovered by {\it Swift/BAT} on 24 August, 2023 \cite[]{Kennea-etal2023,Negoro-etal2023}. Immediate monitoring of the source with {\it MAXI/GSC} in $2-20$ keV reveals an `unusual' peak in its X-ray flux from $150$ mCrab to $3$ Crab within a day \citep{Nakajima-etal2023} that ultimately reaches to peak value $\sim 7$ Crab. Meanwhile, the radio counterpart of the source appears to be consistent with its optical position, and {\it VLA} ($5.25$ GHz) and {\it ATA} ($5$ GHz) independently observe an increase of radio flux from $\sim 18$ to $107$ mJy \cite[]{Miller-Jones-etal2023,Bright-etal2023} just after six days of the source discovery. Further, {\it VLITE} continuously monitor the source in the radio frequency at $338$ MHz \cite[]{Peters-etal2023}. Interestingly, during the fast rising period of the ongoing outburst, strong QPOs in the frequency range $0.44-0.88$ Hz are observed by both {\it NICER} and {\it Swift/BAT} \citep{Draghis-etal2023, Palmer-etal2023}. {\it AstroSat} also observed this source on 02 September, 2023 and a prominent QPO signature was detected in the power spectra at $1.1$ Hz along with harmonic at $2.0$ Hz \cite[]{Kotoch-etal2023a}. In addition, {\it IXPE} reported the detection of polarized emission in hard intermediate state of the source with polarization degree ${\rm PD} \sim 4.1\% \pm 0.2\%$ and polarization angle ${\rm PA} \sim 2.2^\circ \pm 1.3^\circ$. Based on the polarization results and comparing X-ray flux with known BH-XRBs, the inclination and distance of the source are predicted as $i \sim 30^{\circ}-60^{\circ}$ and $1.5$ kpc \citep{Veledina-etal2023}, respectively. 

In this paper, for the first time to the best of our knowledge, we report the detection of energy resolved LFQPO at frequencies $\nu_{_{\rm QPO}} \sim 1.4-2.6$ Hz in hard X-rays ($\sim 100$ keV) during HIMS of Swift J$1727.8-1613$. Needless to mention that the large effective area of {\it AstroSat/LAXPC} makes it possible to study the variability of LFQPO at high energies. We also present the systematic evolution of LFQPO frequencies using {\it AstroSat} observations (augmented with {\it NICER} and {\it Swift} detections) during the onset phase of the outburst. The hardness intensity variation along with the spectral study suggests that the source evolved from LHS to HSS via intermediate states (HIMS and SIMS). At present, the source transits through the decay phase.

The Letter is organized as follows. In \S2, we mention observation details and data analysis procedures for each instruments. In \S3, we present results associated with source outburst, detection and evolution of LFQPO at hard X-rays including spectral modelling. Finally, in \S4, we conclude with discussion.

\begin{figure*}
	\begin{center}
		\includegraphics[width=\textwidth]{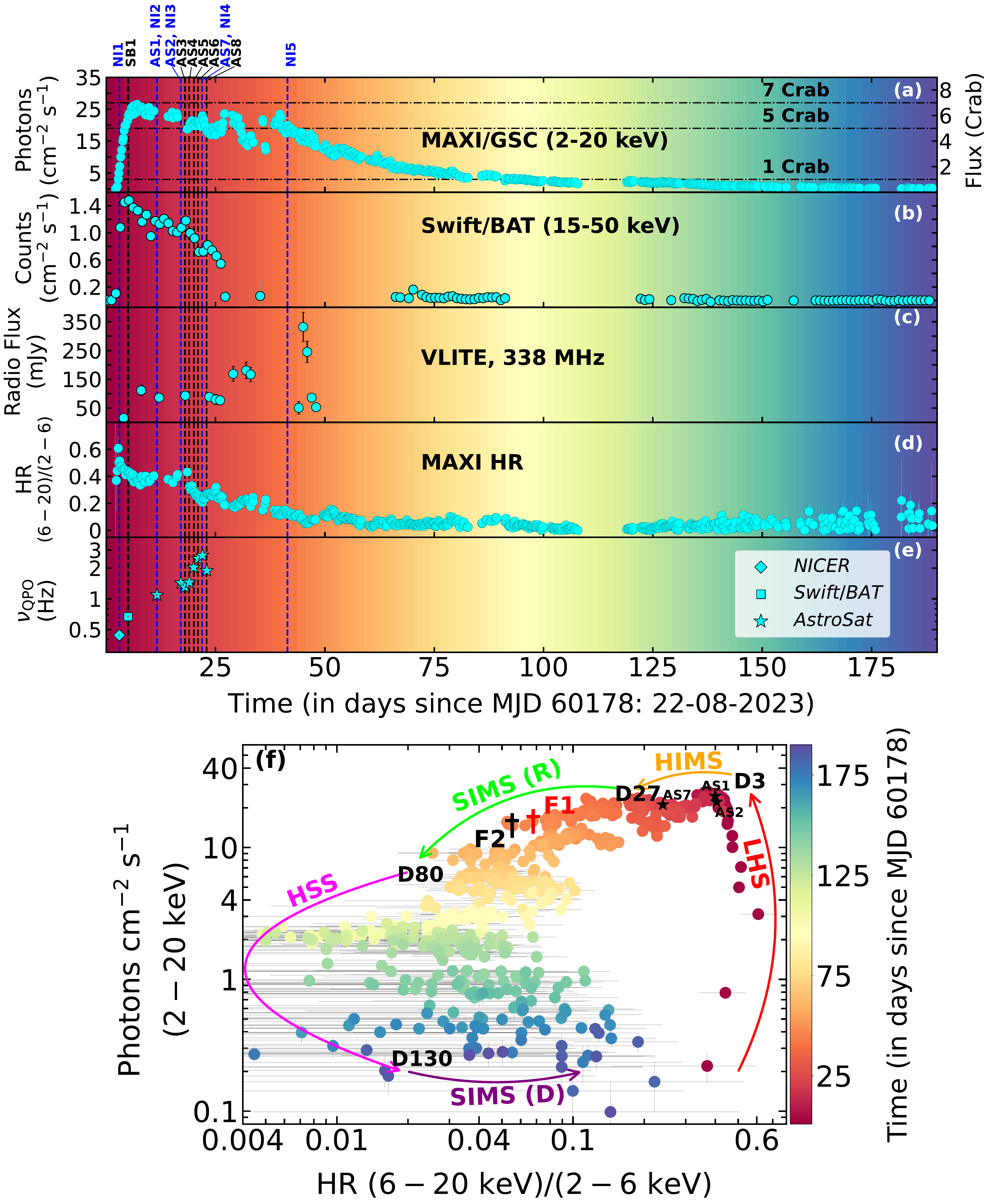}
	\end{center}
	\caption{Long term monitoring of Swift J1727.8$-$1613 with multiple instruments in different energy bands. In panel (a) and (b), variation of {\it MAXI/GSC} and {\it Swift/BAT} count rates in $2-20$ keV and $15-50$ keV energy ranges are shown. Panel (c) represents radio flux with {\it VLITE} at $338$ MHz. Variation of hardness ratio obtained from {\it MAXI/GSC} is plotted in panel (d). Evolution of $\nu_{_{\rm QPO}}$ in Swift J1727.8$-$1613 detected from different instruments is shown in panel (e). Hardness intensity diagram obtained from {\it MAXI/GSC} observations in $2-20$ keV energy band is shown in panel (f). Colorbar indicates day number since source discovery.
	}
	\label{fig:maxi}
\end{figure*}

\section{Observation and Data reduction}

{\it AstroSat} \cite[]{Agrawal-etal2017} observed Swift J1727.8$-$1613 on 02 September, 2023 (MJD 60189) for the first time during a slew mode operation with an offset of $\sim 0.9$ degree from the pointing (RA $= 261.13^{\circ}$, DEC $= -16.01^{\circ}$) for $11$ sec ($T_{\rm start}$ = 14$:$18 UT) exposure only. The source was further observed by {\it AstroSat} as a part of Target of Opportunity (ToO)\footnote{\url{https://astrobrowse.issdc.gov.in/astro\_archive/archive/Home.jsp}} campaign during 08 September, 2023 (MJD 60195) to 14 September, 2023 (MJD 60201) for a total exposure of $\sim 207$ ksec. During slew mode, {\it LAXPC20} was operational although it was switched off during ToO observations for safety reasons to avoid detector saturation due to very high source count rates. {\it LAXPC10} was operational in a low gain mode for ToO observations. Note that we use {\it LAXPC10} data for temporal analysis only, where data was collected in Event Mode (EA).

{\it LAXPC} level-1 data is processed by \texttt{LaxpcSoftv3.4.4}\footnote{\url{https://www.tifr.res.in/~astrosat_laxpc/LaxpcSoft.html}} to extract events, lightcurves, spectra and background files for the good time intervals (GTIs). The {\it LAXPC} background is estimated from the blank sky observations \cite[]{Antia-etal2017,Antia-etal2021,Antia-etal2022} made close to the source observation. While using data from slew mode operation, we use the off-axis response matrix file for {\it LAXPC20} (\texttt{lx20cshm13off50v1.0.rmf}$^2$) for channel to energy conversion to carry out both timing and spectral analyses \cite[]{Antia-etal2017,Baby-etal2021,Katoch-etal2021,Bhuvana-etal2023}. As {\it LAXPC10} operates at low gain mode, the channel to energy mapping remains uncertain. Hence, we obtain energy-channel relation using the $60$ keV calibration peak in the veto anode A8 \cite[]{Antia-etal2017} and a feature in background spectrum (observed before ToO observation) around $30$ keV. These peaks were attributed around channels $70$ and $35$, respectively, in $1024$ channel space of {\it LAXPC10}, suggesting an approximately linear response. Employing this scaling, we estimate the low energy threshold to be around $20$ keV. With this, we adopt a linear relation to obtain energy-resolved lightcurve.	Lightcurves are generated using single event, all layers data of {\it LAXPC10} \cite[]{Sreehari-etal2020,Katoch-etal2021}.

Swift J$1727.8-1613$ is also observed with {\it NICER} almost on a daily basis since its discovery. In this work, we analyze the {\it NICER} observations on 02, 08, and 13 September, 2023 quasi-simultaneous with {\it AstroSat} observations. Two more observations on 25 August and 02 October, 2023 in different phases of the outburst are also analysed. The data is processed using the {\it NICER} data analysis software (\texttt{NICERDAS v10}) available in \texttt{HEASOFT V6.32.1}\footnote{\url{https://heasarc.gsfc.nasa.gov/docs/software/heasoft/}} with the appropriate calibration database. The task \texttt{nicerl2} is used to generate clean event files considering all the standard calibration and data screening criteria. Further, the spectral products are extracted employing \texttt{nicerl3-spect} tool. We select the background model \texttt{3c50} using the flag \texttt{bkgmodeltype=3c50} during the extraction of spectral products.

\section{Modeling \& Results}

\subsection{Outburst Profile and HID}

We study the outburst profile of Swift J1727.8$-$1613 using data from multiple instruments ({\it MAXI}, {\it BAT}, {\it VLITE}) in different energy bands. Since its detection on 24 August 2023, the source seems to exhibit a canonical outburst profile with a sudden rise from quiescence followed by a slow decay. In Fig. \ref{fig:maxi}a-c, we present the results obtained from (a) $0.2$ day averaged monitoring with {\it MAXI/GSC} ($2-20$ keV), (b) $1$ day binned light curve from {\it Swift/BAT} ($15-50$ keV), and (c) radio detections using {\it VLITE} (at $338$ MHz), respectively. During the initial phase of the outburst ($\sim 27$ days), source is observed with `compact' radio emission of flux $\sim 80$ mJy. However, a significant increase of radio emission {$\sim 150$ mJy is observed during the transition from HIMS to SIMS, which is followed by strong radio flare with 
flux $\sim 350$ mJy after $\sim 15$ days \cite[]{Peters-etal2023}. In panel (d), we depict the variation of hardness ratio ($HR$) defined as the ratio of photon counts in $6-20$ keV to $2-6$ keV energy bands of {\it MAXI/GSC}. We show the evolution of detected Type-C QPO frequency ($\nu_{_{\rm QPO}}$) during the onset phase of outburst observed with {\it NICER}, {\it Swift} and {\it AstroSat} in panel (e).

In Fig. \ref{fig:maxi}f, we present the hardness intensity diagram (HID) obtained from the {\it MAXI/GSC} monitoring, where the variation of intensity (photons cm$^{-2}$ s$^{-1}$) in $2-20$ keV energy range is plotted with HR. The obtained results are plotted using color coded filled circles where colorbar indicates the day number since the discovery of the source (see also panels Fig. \ref{fig:maxi}a-d). As the outburst progresses, the source traces different spectral states, namely LHS with HR $\sim 0.6-0.47$ (D0-D3), HIMS with HR $\sim 0.47-0.19$ (D3$-$D27), SIMS(R) with HR $\sim 0.19-0.02$ (D27$-$D80), a long-lived HSS with HR $\lesssim$ 0.02 (D80$-$D130), respectively. At present, the source probably evolves through the decay phase of SIMS (D130$-$till date) with HR $\gtrsim 0.02$ and flux lavel $\sim70$ mCrab of the ongoing outburst. Note that the observed radio detection which is strongly correlated with the spectral states, further confirms the transition from HIMS to SIMS ($\sim {\rm D27}$) of the source along with strong radio flare detected during SIMS. 

\subsection{Detection and Evolution of LFQPO in hard X-rays}

\begin{figure}
	\begin{center}
		\includegraphics[width=\columnwidth]{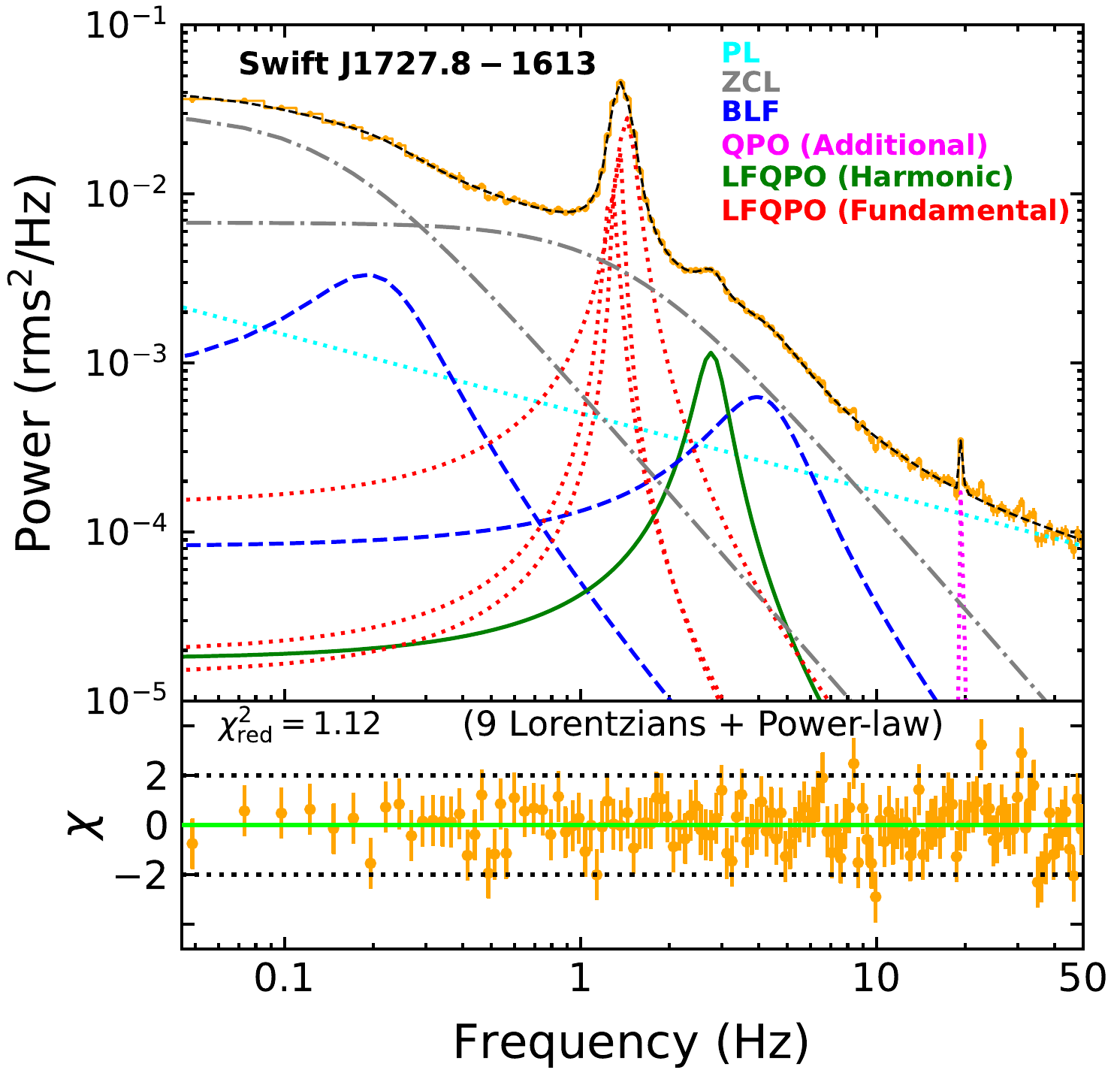}
	\end{center}
	\caption{Best fitted PDS of AS2 observation (MJD 60195) of Swift J$1727.8-1613$ in $20-100$ keV energy band. The different model components (nine \texttt{Lorentzian} and one \texttt{power-law}) corresponding to different characteristics are used to fit the entire PDS. The variation of residuals is shown in the bottom panel. See the text for details.
	}
	\label{fig:pdsAS2}
\end{figure}

We examine the power density spectra (PDS) using the archival {\it LAXPC10} data during the outburst phase (see \S2, Fig. \ref{fig:maxi}a-b). While doing so, $0.01$ s time binned lightcurves in $20-100$ keV energy range are used to generate PDS up to the Nyquist frequency ($50$ Hz) using ftool \texttt{powspec}\footnote{\url{https://heasarc.gsfc.nasa.gov/xanadu/xronos/examples/powspec.html}} within \texttt{HEASOFT V6.32.1}\footnote{\url{https://heasarc.gsfc.nasa.gov/lheasoft/release_notes.html}}. We choose $4096$ newbins per interval that results 2048 frequency points in the unbinned PDS. These 2048 frequency points are further geometrically rebinned with a factor of 1.03 to obtain the resultant PDS with 140 frequency bins in unit of $\rm rms^{2}/Hz$ \cite[]{Sreehari-etal2019}. Accordingly, each frequency bin contains $\sim 15$ data points which are averaged out. Because of this, the underlying likelihood function tends to follow a Gaussian distribution \cite[]{vanderKlis-1989,Vaughan2010} in accordance with the central limit theorem and therefore, the chi-square statistics remain valid for PDS modelling \cite[]{Papadakis-etal1993}. 

Each PDS is modelled with a combination of multiple \texttt{Lorentzian} \cite[]{van-der-Klis-1994a,van-der-Klis1994b,Nowak-etal2000,Belloni-etal2002,Belloni-etal2005,Kushwaha-etal2021} and a \texttt{power-law} component within \texttt{XSPEC} environment. A \texttt{Lorentzian} is characterized by three parameters: centroid (LC), width (LW) and normalization (LN). In Fig. \ref{fig:pdsAS2}, we present the various model components to describe the PDS continuum and QPO features of epoch AS2 (MJD 60195) in $20-100$ keV energy range. While doing so, we use two zero-centred \texttt{Lorentzian} (ZCL; dot-dashed in grey), two bump-like features (BLF; dashed in blue) at $\sim 0.2$ Hz and $\sim 4$ Hz and one power-law component (PL; dotted in cyan) to model the entire continuum in $0.1-50$ Hz frequency range. In addition, in modelling the complex nature of LFQPO feature \cite[see also][]{Dotani-etal1989,Belloni-etal2002}, we use three \texttt{Lorentzian} (dotted in red) at three distinct nearby frequencies of $\sim 1.27$ Hz, $\sim 1.34$ Hz and $\sim 1.43$ Hz. Further, one additional \texttt{Lorentzian} (solid in green) profile is required to model the harmonic feature at $\sim 2.76$ Hz. Finally, one more \texttt{Lorentzian} (dotted in magenta) is needed to model a possible QPO-like feature at $\sim 20$ Hz. With this, we obtain best fit PDS with a $\chi_{\rm red}^{2}$ ($\chi^2/d.o.f$) of $131/117=1.12$ that yields a strong LFQPO feature of frequency $\nu_{_{\rm QPO}} \sim 1.43$ Hz in $20-100$ keV energy band. We present the best fitted residuals variation in the bottom panel of Fig. \ref{fig:pdsAS2}. Following the above approach, we further carry out the modelling of the energy dependent PDS of epoch AS2 (MJD 60195). The best fitted model parameters associated to individual model components used in the PDS fitting are tabulated in Table \ref{table:PDS_parameters}. Similar model combinations are also used to fit the PDS ($20-100$ keV) of epoch AS7 (MJD 60199) with the exception that the bump-like features ($\sim 1$ Hz, $\sim 5$ Hz and $\sim 8$ Hz) and complex LFQPO features ($\sim 2.17$ Hz, $\sim 2.66$ Hz and $\sim 3.29$ Hz) are present at different frequencies.
	
We verify the significance of the LFQPO features by comparing the chi-square statistics obtained from the best fit due to the inclusion of respective \texttt{Lorentzian} components. We also examine $\chi^2$ statistics of the PDS modelling with and without QPO feature. Towards this, we compute the change in fitted chi-square value per degrees of freedom (defined as $\Delta \chi_{\rm dof}^{2}= (\chi_{b}^{2} - \chi_{a}^{2})/\Delta {\rm dof}$) using best fitted chi-square value before ($\chi_{b}^{2}$) and after ($\chi_{a}^{2}$) the inclusion of respective \texttt{Lorentzian} component corresponding to the difference in degrees of freedom ($\Delta {\rm dof}$). Accordingly, the best fit statistics is realized with the improvement of $\Delta \chi_{\rm dof}^{2}$ due to the inclusion of respective \texttt{Lorentzian} components, which are presented in Table \ref{table:PDS_parameters}. As an example, in $20-100$ keV, $\Delta \chi_{\rm dof}^{2}$ are computed due to the difference of best fitted PDS for L1-L8 components and L1-L9 components, which renders $\Delta \chi_{\rm dof}^{2} = 309/3=103$. Similar approach is followed in determining the chi-square statistics for other energy bands. The estimated $\Delta \chi_{\rm dof}^{2}$ due to the inclusion of respective \texttt{Lorentzian} components are presented in Table \ref{table:PDS_parameters} for all PDS of AS2. Needless to mention that the requirement of the different model components to obtain the best fit PDS is clearly justified as $\Delta \chi_{\rm dof}^{2}$ improves significantly. We continue to carry out the analyses for AS7 as well, however we refrain in presenting it to avoid repetition.

Following the standard approach \cite[and references therein]{Belloni-etal2013,Sreehari-etal2019,Majumder-etal2022,Majumder-etal2023}, we estimate the {\it significance} ($\sigma=LN/{\rm err}_{\rm neg} $), and {\it Q-factor} ($\equiv \nu_{\rm QPO}/\Delta \nu$) for QPO features. We obtain $\sigma$ and {\it Q-factor} for fundamental LFQPO as $5.46~\sigma$ ($22.97~\sigma$) and $7.14$ ($5.02$) for epoch AS2 (AS7), respectively. We further estimate the percentage rms amplitude \cite[$rms_{_{\rm QPO}}$;][]{Ribeiro-etal2019,Sreehari-etal2020} of the detected LFQPO and obtained as $10.95\%$ ($9.21\%$) for epoch AS2 (AS7). Note that in both AS2 and AS7 epochs, we consider the centroid frequency of LFQPO that renders maximum power. Subsequently, the LFQPO parameters are computed for this component only. All the best fit model and estimated parameters are tabulated in Table \ref{table:pds_fit}. In Fig. \ref{fig:maxi}e, we present the evolution of $\nu_{_{\rm QPO}}$ ($1.09-2.66$ Hz) during the entire {\it AstroSat} campaign along with {\it NICER} and {\it Swift/BAT} detection in $0.44-0.8$ Hz frequency range.
	
\begin{figure}
	\begin{center}
		\includegraphics[width=\columnwidth]{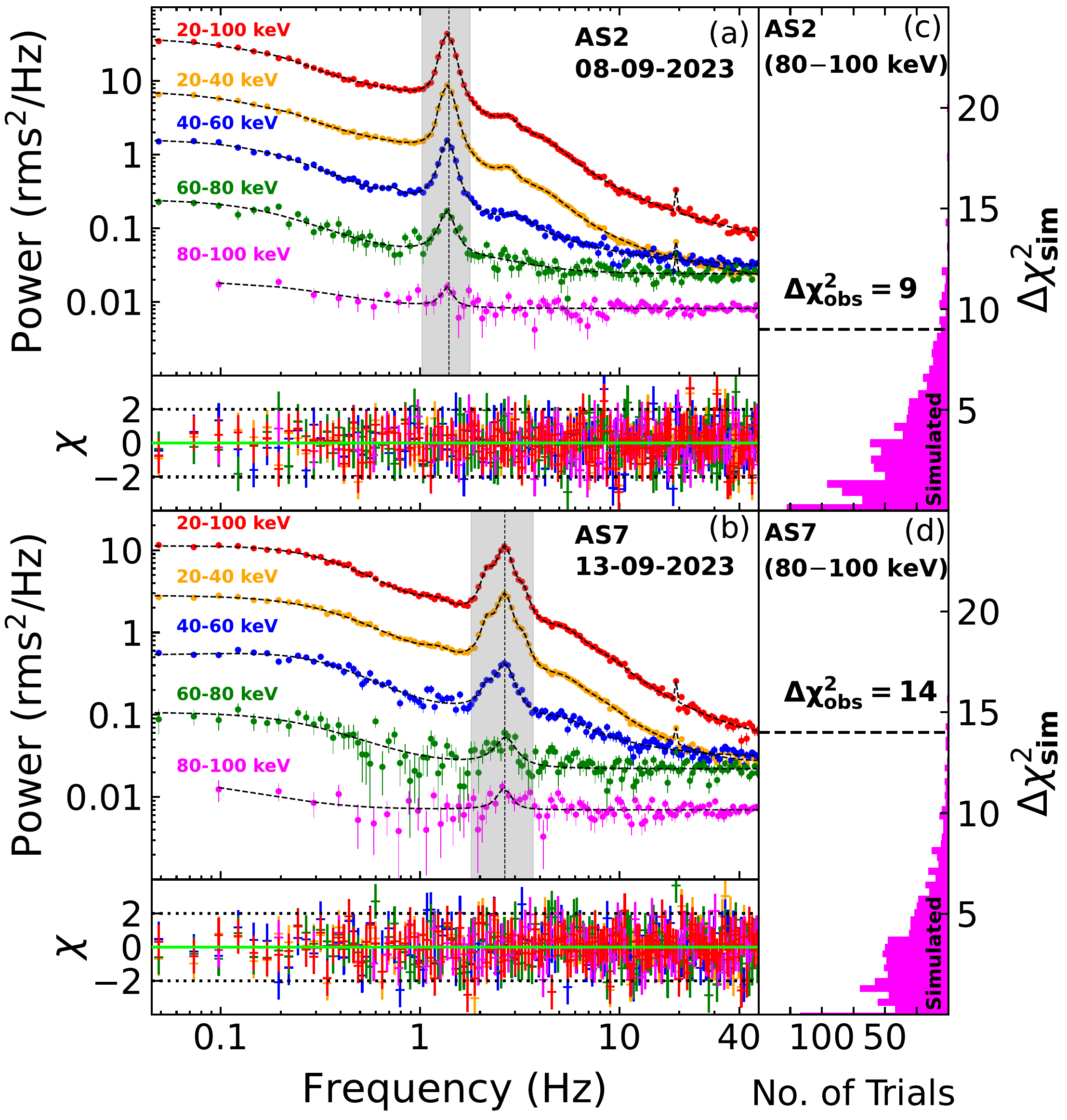}
	\end{center}
	\caption{Best fitted power spectra of Swift J$1727.8-1613$ in different energy bands (obtained from {\it AstroSat/LAXPC10} observations). Panels (a) and (b) are for epoch AS$2$ and AS$7$ and residual variations are shown at the bottom of each panel. For clarity, power spectra in $20-100$ keV, $20-40$ keV, $40-60$ keV, $60-80$ keV and $80-100$ keV energy bands are scaled by multiplying with constant $950$, $150$, $40$, $20$, and $5$. Panels (c) and (d) show the results obtained from \texttt{simftest} in $80-100$ keV energy band of both observations. See the text for details.
	}
	\label{fig:pds}
\end{figure}

We further study the energy dependent properties of the detected LFQPO features. Towards this, we generate PDS in different energy bands, namely $20-40$ keV, $40-60$ keV, $60-80$ keV and $80-100$ keV. We detect strong Type-C LFQPO features in all the aforementioned energy bands (see Table \ref{table:pds_fit}) and observe that the centroid frequency ($\nu_{_{\rm QPO}}$) remains independent of X-ray photon energy as shown in Fig. \ref{fig:pds} (AS2 and AS7 epochs). It may be noted that a ZCL component along with an additional \texttt{Lorentzian} at $\nu_{\rm QPO}$ are used to fit the PDS in higher energy band ($80-100$ keV). In order to confirm the detection significance of QPO, we perform simulation using \texttt{simftest} inside \texttt{XPSEC} \cite[]{Athulya-etal2022,Bhuvana-etal2023,Sharma-etal2023,Li-etal2023} to obtain the probability of best fit without the LFQPO features in $80-100$ keV energy band. In Fig. \ref{fig:pds}(c-d), we present the distribution of the difference of chi-square values with and without LFQPO feature for $1000$ simulated power spectra. The change of the chi-square value obtained from the real observed data ($\Delta \chi_{\rm obs}^{2}$) by modelling the LFQPO feature is found outside the distribution of the change in chi-square value ($\Delta \chi_{\rm sim}^{2}$) obtained from the simulation. These findings evidently indicate that the detected $\nu_{_{\rm QPO}}$ is directly associated with the hard X-ray emissions. The energy dependent model fitted and estimated parameters along with fit statistics are tabulated in Table \ref{table:pds_fit}.

\begin{figure}
	\begin{center}
		\includegraphics[width=0.9\columnwidth]{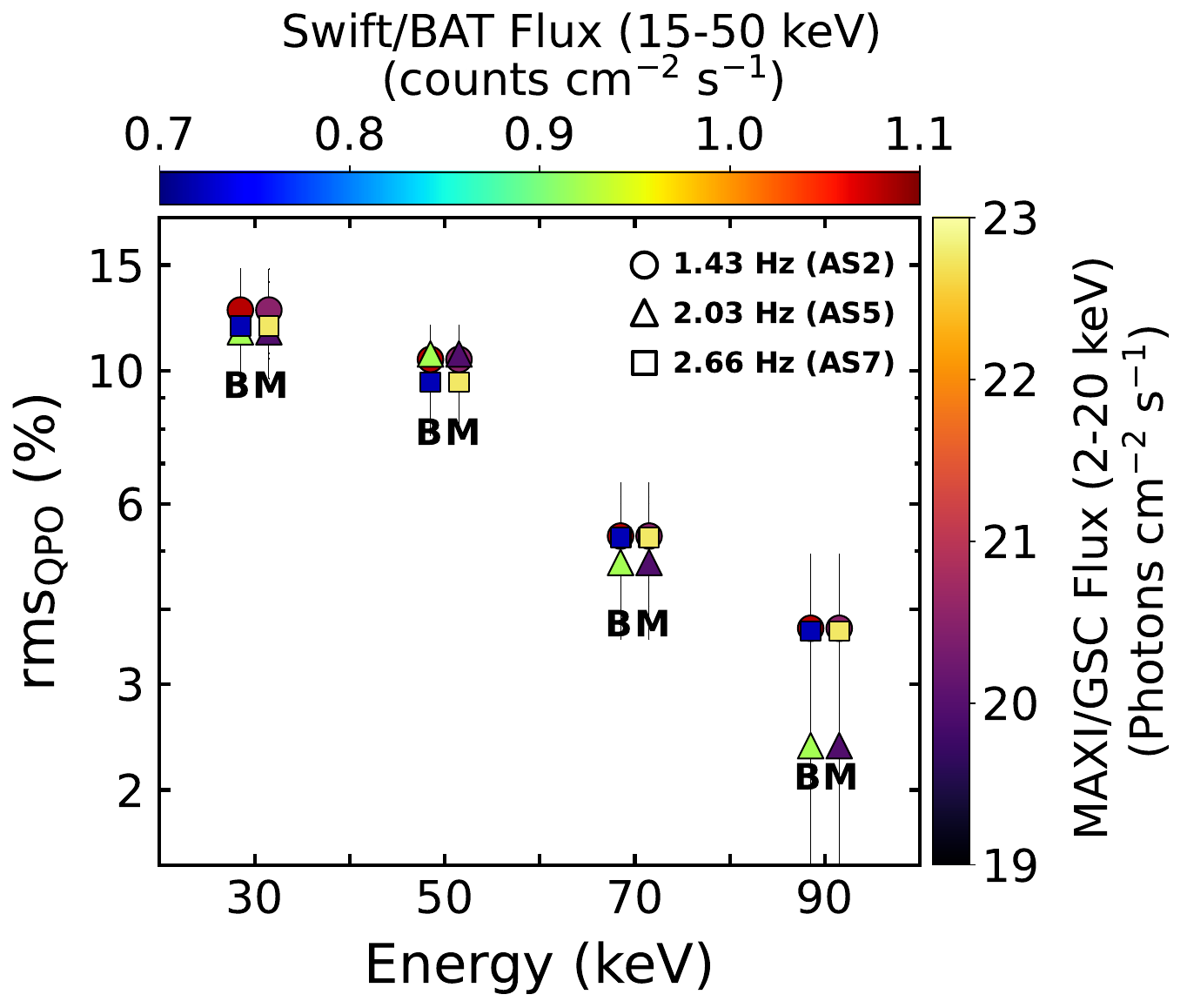}
	\end{center}
	\vskip -0.5 cm
	\caption{Plot of ${\rm rms}_{_{\rm QPO}}$ with energy visualized in terms of {\it SWIFT/BAT} (B) and {\it MAXI/GSC} (M) fluxes. Color coded filled circles, triangles and squares denote frequency of LFQPO of $\nu_{_{\rm QPO}}=1.43$ Hz (AS2), $2.03$ Hz (AS5) and $2.66$ Hz (AS7), respectively. Points marked with `B' and `M' are horizontally shifted for better visibility. Horizontal and vertical colorbars denote {\it Swift/BAT} and {\it MAXI/GSC} fluxes.
	}
	\label{fig:corr_plt}
\end{figure}

Furthermore, we study the variation of energy-resolved rms amplitude ($rms_{_{\rm QPO}}$) of the LFQPOs as function of {\it SWIFT/BAT} (B) and {\it MAXI/GSC} (M)  fluxes. The obtained results are shown in Fig. \ref{fig:corr_plt}, where filled colored circles, triangle and squares denote the Type-C LFQPO of $\nu_{_{\rm QPO}} = 1.43$ Hz (AS2), $2.03$ Hz (AS5) and $2.66$ Hz (AS7) observed with {\it AstroSat} in different energy bands. In the figure, all points marked with `B' and `M' are shifted horizontally for better clarity. The horizontal and vertical colorbars demonstrate the quasi-simultaneous {\it SWIFT/BAT} (counts in $\rm cm^{-2}$ $\rm s^{-1}$ within $15-50$ keV) and {\it MAXI/GSC} (photons in $\rm cm^{-2}$ $\rm s^{-1}$ within $2-20$ keV) fluxes. We observe $rms_{_{\rm QPO}} \sim 13\%$ at $20-40$ keV which drops down around $\sim 2.5\%$ at $80-100$ keV for all epochs under considerations. It is noteworthy that {\it SWIFT/BAT} flux anti-correlates with $\nu_{_{\rm QPO}}$ in all energy band, whereas {\it MAXI/GSC} flux increases with $\nu_{_{\rm QPO}}$. 

\begin{figure}
	\begin{center}
		\includegraphics[width=\columnwidth]{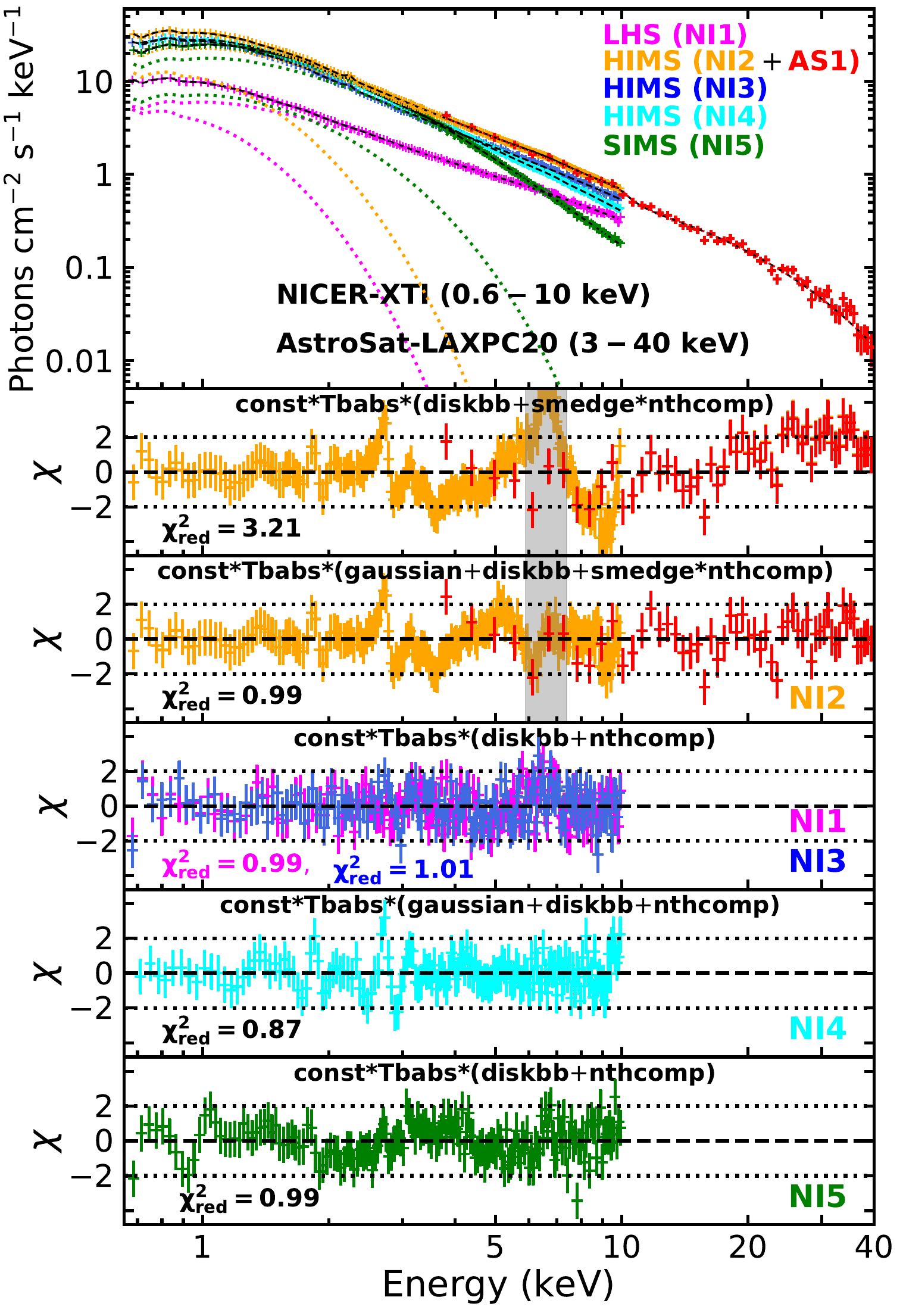}
	\end{center}
	\vskip -0.5 cm
	\caption{{\it Upper panel}: Best fit energy spectra of Swift J1727.8$-$1613 with {\it NICER} and {\it AstroSat} observations during the onset phase of the outburst. {\it Lower panels}: Residuals variations obtained from the fitting with different model combinations. 
		}
	\label{fig:spectra}
\end{figure}

\subsection{Spectral Energy Distribution}

We investigate the spectral properties of Swift J$1727.8-1613$ using combined {\it NICER/XTI} and {\it AstroSat/LAXPC20} observations in $0.6-40$ keV energy range. The spectral analysis is carried out for quasi-simultaneous {\it NICER} and {\it AstroSat} observations during epoch AS1 and NI2. In addition, the spectral distribution of the source is studied using {\it NICER} observations NI1, NI3, NI4 and NI5 in $0.6-10$ keV energy range, when {\it AstroSat/LAXPC20} monitoring were not available.
	
Spectra are modelled using \texttt{XSPEC V12.13.Oc} in \texttt{HEASOFT V6.31.1}. We adopt a model combination \texttt{const*Tbabs*(gaussian+diskbb+smedge*nthcomp)} to describe the spectral distribution. Here, \texttt{const} takes care of the cross calibration between the spectra of different instruments and \texttt{Tbabs} is used for the galactic absorption. The models \texttt{diskbb} \citep{Makishima-etal1986} and \texttt{nthcomp} \citep{Zdziarski-etal1996} represent the standard accretion disc and thermal Comptonization components. The \texttt{gaussian} is incorporated in the spectral fitting to model the iron line emission at $\sim 6.4$ keV. Additionally, a \texttt{smedge} component is used at $\sim 9.94$ keV to improve the residual variations only for AS1+NI2 observations. It may be noted that {\it NICER} spectra showed two instrument lines at $1.8$ keV and $2.2$ keV, which are modelled using two additional \texttt{gaussian}. With this, we obtain an acceptable fit with $\chi^2_{\rm red}$ of $0.99$. The resultant fit indicates the presence of a weak disc signature having temperature of $0.36 \pm 0.01$ keV. Moreover, a hard Comptonized spectral tail of photon index $1.84 \pm 0.01$ and electron temperature $6.83_{-0.29}^{+0.30}$ keV are obtained. We compute the flux associated with different spectral components as well as total bolometric flux using the convolution model \texttt{cflux} in $0.5-50$ keV energy range. Following \cite{Zdziarski-etal1996,Majumder-etal2022}, we estimate the optical depth ($\tau$) and Compton y-parameter (y-par). We find that the {\it NICER} spectra are well described with the aforementioned model without the \texttt{smedge} component. All the model fitted and estimated spectral parameters are tabulated in Table \ref{tab:spectra}. The best fit spectra obtained from {\it NICER} and {\it AstroSat} observations and the corresponding residuals are shown in Fig. \ref{fig:spectra}.

\section{Discussion}

In this paper, we report the discovery of evolving LFQPO features ($1.4 - 2.6$ Hz) at hard X-rays ($\sim 100$ keV) with {\it AstroSat} during the `unusual' outburst phase of Swift J1727.8$-$1613. The source traces the canonical hysteresis loop in the HID with a fugacious HSS followed by a decay phase. During the {\it AstroSat} campaign, the source was in HIMS dominated by the Comptonized emission ($\sim 83-88\%$) with weak signature of disk emission ($\sim 9-16\%$) and temperature $kT_{\rm in} \sim 0.4-0.6$ keV.

The most fascinating results we report in this work are the significant detection of energy dependent LFQPO features above $40$ keV, particularly in the energy range of  $80-100$ keV with $Q> 6$ and $\sigma > 3$ (see Table \ref{table:pds_fit} and Fig. \ref{fig:pds}), when the source was in HIMS. We observe that as the outburst progresses, $\nu_{_{\rm QPO}}$ is evolved from $\sim 0.44$ Hz to $2.66$ Hz during LHS ({\it NICER} and {\it Swift/BAT}) and HIMS ({\it AstroSat}) (see Fig. \ref{fig:maxi}). The unfolded energy spectra (Fig. \ref{fig:spectra}) clearly indicate that the spectral energy distribution is dominated by non-thermal emission with electron temperature $kT_e \sim 4-7$ keV and spectral index $\Gamma \sim 1.6 -1.9$ (see Table 2).

The detection of LFQPO ($\nu_{_{\rm QPO}}<1$ Hz) at high energies ($>$100 keV) was earlier reported for few BH-XRBs (MAXI J1820$+$070, \cite{Ma-etal2021} and MAXI J1803$-$298, \cite{Wang-etal2021}), however strong LFQPO ($\nu_{_{\rm QPO}}\sim 1-10$ Hz) at energies $<$100 keV were observed in MAXI J1535$-$571 \cite[]{Huang-etal2018} and MAXI J1631$-$479 \cite[]{Bu-etal2021} sources. It was suggested that the Lense-Thirring (LT) precession of the inner hot flow can account for the possible origin of Type-C LFQPO of BH-XRBs \cite[see][for details]{Ingram-etal2016}. Considering this, efforts were further given to investigate the Type-C LFQPOs using LT precession of a small scale jet, where jet rotates and twists around the spin axis of BH resulting the modulation of the observed flux \cite[]{Ma-etal2021}. However, this scenario bears limitations as it requires steady increase of jet precession in order to explain the evolution of LFQPO of $\nu_{_{\rm QPO}} \sim 1.09 -2.66$ Hz as detected in Swift J1727.8$-$1613 at high energies $> 40$ keV. In addition, while explaining the evolution of $\nu_{\rm QPO}$ considering the jet precession model for low inclination source ($\sim 29^{\circ}$), \cite{Bu-etal2021} claimed that the frequency of Type-C LFQPO generally increases with the decrease of jet height. However, jets are unlikely to be decoupled in LHS/HIMS \cite[]{Fender-etal2004,Fender-etal2009} and hence, this model also suffers shortcomings to explain the evolution of QPO for Swift J1727.8$-1613$, although the orbital inclination of this source is recently reported as $40^\circ$ \cite[]{Peng-etal2024}. Moreover, as there is no energy dependence of $\nu_{\rm QPO}$ observed for this source (see Table \ref{table:pds_fit}), it is implausible that any differential precession is involved in exhibiting the LFQPO \cite[]{Ingram-etal2016}. Needless to mention that the evolution of the Type-C LFQPO ($\nu_{\rm QPO} \sim 0.1-30$ Hz) is observed in the BH-XRBs having time scale of $5-20$ days \cite[]{Remillard-McClintock2006,Nandi-etal2012,Radhika-Nandi2014,Radhika-etal2016}, which remains challenging to explain till date.

What is more is that the Type-B QPOs are generally observed during the spectral state transition from HIMS to SIMS as well as in SIMS \cite[][and references therein]{Fender-etal2004,Fender-etal2009,Radhika-etal2016,Huang-etal2018,Ingram-Motta2019}. However, for Swift J1727.8$-1613$, we find that all the LFQPOs ($ \nu_{\rm QPO} \sim 0.44-2.66$ Hz) are observed in LHS and HIMS with $rms_{_{\rm QPO}}\sim 10\%$ and $Q > 5$, which are indeed the prime characteristics of a Typc-C QPO \cite[]{Belloni-etal2005,Nandi-etal2012,Ingram-Motta2019}.

Meanwhile, alternate possibilities are also suggested that comprehend the local inhomogeneity in accretion disk yielded the time varying modulation of the inner `hot' flow and hence, it contributes in resulting $\nu_{_{\rm QPO}}$ \cite[]{Huang-etal2018,Bu-etal2021}. In reality, this conjecture resembles the corona oscillation scenario caused due to the undulation of the `hot' and `dense' down-stream flow in the vicinity of the black holes \cite[]{Molteni-etal1996}. Since $\nu_{_{\rm QPO}}$ is comparable to the inverse of the infall time scale ($t_{\rm infall}$) \cite[]{Chakrabarti-Manickam2000} and $t_{\rm infall}$ strongly depends on corona geometry, it is more likely that $\nu_{_{\rm QPO}}$ increases as the region occupied by the down-stream flow decreases \cite[]{Chakrabarti-etal2008,Nandi-etal2012,Iyer-etal2015}. Indeed, the dynamics of the down-stream flow can be regulated by tuning the accretion parameters, namely accretion rate and viscosity \cite[]{Chakrabarti-Molteni1993,Das-etal2014}. When the soft photons from the upstream are up-scattered at the inner `hot' flow, high energy X-ray emissions are produced because of the inverse-Comptonization process \cite[]{Chakrabarti-Titarchuk1995,Mandal-Chakrabarti2005}. Accordingly, LFQPO is likely to observe at high energies whenever the source remains in harder spectral state dominated by the Comptonized emission in presence of `weak' disc signature.

To summarize, we report the first detection of evolving LFQPO features of Swift J1727.8$-$1613 above $40$ keV in HIMS observed with {\it AstroSat}. In addition, to the best of our knowledge, Swift J1727.8$-$1613 is the third known BH-XRB that exhibits significant detection of LFQPO ($>1$ Hz) at higher energies ($\sim 100$ keV).

\section*{Acknowledgments}

Authors thank the anonymous reviewer for constructive comments and useful suggestions that help to improve the quality of the manuscript. AN thanks GH, SAG; DD, PDMSA; Associate Director and Director, URSC for encouragement and continuous support to carry out this research. SD thanks Science and Engineering Research Board (SERB) of India for support under grant MTR/2020/000331. This publication uses the data from the {\it AstroSat} mission of the Indian Space Research Organisation (ISRO), archived at the Indian Space Science Data Centre (ISSDC). This work has used the data from the {\it LAXPC} Instruments developed at TIFR, Mumbai, and the LAXPC-POC at TIFR is thanked for verifying and releasing the data. We also thank the {\it AstroSat} Science Support Cell hosted by IUCAA and TIFR for providing the \texttt{LAXPCSOFT} software which we used for {\it LAXPC} data analysis. This work has also use data from Monitor of All-sky X-ray Image ({\it MAXI}) data provided by Institute of Physical and Chemical Research (RIKEN), Japan Aerospace Exploration Agency (JAXA), and the {\it MAXI} team.  Also this research made use of software provided by the High Energy Astrophysics Science Archive Research Center (HEASARC) and NASA's Astrophysics Data System Bibliographic Services. This publication further uses the data from the {\it NICER} missions, archived at the HEASARC data centre.

\section*{Data Availability}

Data used for this publication are currently available at the Astrobrowse (AstroSat archive) website (\url{https://astrobrowse.issdc.gov.in/astro\_archive/archive}) of the Indian Space Science Data Center (ISSDC).

\begin{landscape}
	\begin{table}
		\centering
		\caption{All the model parameters obtained from the best fitted PDS of epoch AS2 (MJD 60195) in different energy bands. Here, $\alpha_{\rm PL}$, $norm_{\rm PL}$ and $L_{i} (i=1,2,3,4,5,6,7,8,9)$ represent the power-law index, normalization and \texttt{Lorentzian} components used in the fitting, respectively. LC LW and LN are the \texttt{Lorentzian} centroid frequency, FWHM and normalization, respectively. A \texttt{constant} component (instead of \texttt{power-law}) is used for the modeling of $60-80$ keV and $80-100$ keV energy band PDS. $\Delta \chi_{\rm dof}^{2}$ represents the improvement in $\chi^{2}$ per degrees of freedom due to the inclusion of respective model components. Parameters in bold font indicate the fundamental QPO components. See the text for details.}
		
		\renewcommand{\arraystretch}{1.5}
		\resizebox{1.34\textwidth}{!}{%

		\begin{tabular}{c @{\hspace{0.3cm}} c c @{\hspace{0.3cm}} c @{\hspace{0.3cm}} c @{\hspace{0.3cm}} c @{\hspace{0.3cm}} c @{\hspace{0.3cm}} c @{\hspace{0.3cm}} c @{\hspace{0.3cm}} c @{\hspace{0.3cm}} c @{\hspace{0.3cm}} c @{\hspace{0.3cm}} c @{\hspace{0.3cm}} c @{\hspace{0.3cm}} c c}
	
	\hline

	& \multicolumn{2}{|c|}{PL} & & & \multicolumn{2}{|c|}{ZCL} & \multicolumn{2}{|c|}{BLF} & QPO (Additional) & LFQPO (Harmonic) & \multicolumn{3}{|c|}{LFQPO (Fundamental)} & \\
	
	\cline{12-14}
	\cline{8-9}
	\cline{6-7}
	\cline{2-3}

	Energy & $\alpha_{\rm PL}$ & $norm_{\rm PL}$ & & & $\rm L_1$ & $\rm L_2$ & $\rm L_3$ & $\rm L_4$ & $\rm L_5$ & $\rm L_6$ & $\rm L_7$ & $\rm L_8$ & $\rm L_9$ & $\chi^2/d.o.f$ \\

	(keV) & & $(\times  10^{-4})$ & & & & & & & & & & & & \\
	
	\hline
	
	& & & & $\Delta \chi_{\rm dof}^{2}$ & $30$ & $18$ & $62$ & $4$ & $46$ & $41$ & $18$ & $12$ & $103$ & \\	\hline	
	
	& & & & LC & 0.0 & 0.0 & $3.94_{-0.15}^{+0.13}$ & $0.19_{-0.05}^{+0.04}$ & $19.46_{-0.06}^{+0.02}$ & $2.76_{-0.02}^{+0.02}$ & $1.27_{-0.01}^{+0.01}$ & $1.34_{-0.01}^{+0.01}$ & $\bf 1.43_{-0.02}^{+0.02}$ & \\
	
	$20-100$ & $0.46_{-0.09}^{+0.08}$ & $5.05_{-0.89}^{+1.02}$ & & LW & $2.88_{-0.38}^{+0.37}$ & $0.30_{-0.05}^{+0.05}$ & $3.05_{-0.30}^{+0.31}$ & $0.2^{\dagger}$ & $0.1^{\dagger}$ & $0.69_{-0.09}^{+0.09}$ & $0.1^{\dagger}$ & $0.06^{\dagger}$ & $\bf 0.20_{-0.01}^{+0.03}$ & $131/117$  \\
	
	& & & & LN & $0.015_{-0.002}^{+0.002}$ & $0.007_{-0.001}^{+0.001}$ & $0.0027_{-0.0004}^{+0.0005}$ & $0.0009_{-0.0003}^{+0.0004}$ & $0.00013_{-0.00001}^{+0.00001}$ & $0.0012_{-0.0002}^{+0.0002}$ & $0.0019_{-0.0003}^{+0.0003}$ & $0.0027_{-0.0006}^{+0.0006}$ & $\bf 0.00915_{-0.00169}^{+0.00071}$ &  \\
	
	\hline
	& & & & $\Delta \chi_{\rm dof}^{2}$ & $54$ & $56$ & $42$ & $3$ & $29$ & $41$ & $43$ & $20$ & $158$ & \\ \hline
	
	& & & & LC & 0.0 & 0.0 & $3.94_{-0.15}^{+0.13}$ & $0.22_{-0.02}^{+0.02}$ & $19.28_{-0.12}^{+0.13}$ & $2.76_{-0.02}^{+0.02}$ & $1.27_{-0.01}^{+0.01}$ & $1.38_{-0.02}^{+0.01}$ & $\bf 1.42_{-0.02}^{+0.02}$ & \\
	
	$20-40$ & $0.31_{-0.05}^{+0.04}$ & $5.00_{-0.75}^{+0.85}$ & & LW & $2.82_{-0.25}^{+0.24}$ & $0.32_{-0.02}^{+0.02}$ & $3.01_{-0.28}^{+0.29}$ & $0.1^{\dagger}$ & $0.33_{-0.16}^{+0.15}$ & $0.72_{-0.09}^{+0.11}$ & $0.07^{\dagger}$ & $0.1^{\dagger}$ & $\bf 0.22_{-0.02}^{+0.02}$ & $128/116$  \\
	
	& & & & LN & $0.019_{-0.001}^{+0.001}$ & $0.0097_{-0.0006}^{+0.0006}$ & $0.0035_{-0.0006}^{+0.0007}$ & $0.00033_{-0.00015}^{+0.00016}$ & $0.00015_{-0.00003}^{+0.00003}$ & $0.0017_{-0.0003}^{+0.0003}$ & $0.0028_{-0.0006}^{+0.0003}$ & $0.0055_{-0.0009}^{+0.0009}$ & $\bf 0.00943_{-0.00145}^{+0.00153}$ & \\ 
	
	\hline
	& & & & $\Delta \chi_{\rm dof}^{2}$ & $105$ & $-$ & $60$ & $9$ & $-$ & $-$ & $-$ & $-$ & $457^{\boxdot}$ & \\ \hline
	
	& & & & LC & 0.0 & $-$ & $2.87_{-0.13}^{+0.10}$ & $0.72_{-0.03}^{+0.03}$ & $-$ & $-$ & $-$ & $-$ & $\bf 1.38_{-0.01}^{+0.01}$ & \\
	
	$40-60$ & $0.21_{-0.04}^{+0.04}$ & $17.61_{-2.23}^{+2.29}$ & & LW & $0.45_{-0.02}^{+0.02}$ & $-$ & $2.76_{-0.43}^{+0.52}$ & $0.28_{-0.12}^{+0.17}$ & $-$ & $-$ & $-$ & $-$ & $\bf 0.23_{-0.02}^{+0.02}$ & $135/127$  \\
	
	& & & & LN & $0.013_{-0.001}^{+0.001}$ & $-$ & $0.007_{-0.001}^{+0.001}$ & $0.0009_{-0.0003}^{+0.0004}$ & $-$ & $-$ & $-$ & $-$ & $\bf 0.01291_{-0.00056}^{+0.00064}$ & \\
	
	
	\hline
	& & \texttt{constant} $(\times  10^{-4})$ & & $\Delta \chi_{\rm dof}^{2}$ & $87$ & $-$ & $9$ & $-$ & $16$ & $-$ & $-$ & $-$ & $38$ & \\ \hline
	
	& & & & LC & 0.0 & $-$ & $1.51_{-0.79}^{+0.81}$ & $-$ & $19.34_{-0.22}^{+0.15}$ & $-$ & $-$ & $-$ & $\bf 1.37_{-0.01}^{+0.01}$ &  \\
	
	$60-80$ & $-$ & $12.02_{-0.24}^{+0.22}$ & & LW & $0.43_{-0.06}^{+0.06}$ & $-$ & $4.18_{-1.27}^{+2.68}$  & $-$ & $0.1^{\dagger}$ & $-$ & $-$ & $-$ & $\bf 0.24_{-0.04}^{+0.04}$ & $153/129$ \\
	
	& & & & LN & $0.0036_{-0.0005}^{+0.0004}$ & $-$ & $0.0033_{-0.0012}^{+0.0013}$ & $-$ & $0.0007_{-0.0001}^{+0.0001}$ & $-$ & $-$ & $-$ & $\bf 0.00243_{-0.00031}^{+0.00040}$ & \\

	\hline
	& & & & $\Delta \chi_{\rm dof}^{2}$ & $10$ &$-$ & $-$& $-$& $-$& $-$& $-$ & $-$ & $3$ & \\ \hline
	
	& & & & LC & 0.0 & $-$ & $-$ & $-$ & $-$ & $-$ & $-$ & $-$ & $\bf 1.37_{-0.05}^{+0.04}$ & \\
	
	$80-100$ & $-$ & $16.31_{-0.26}^{+0.26}$ & & LW & $0.61_{-0.22}^{+0.39}$ & $-$ & $-$ & $-$ & $-$ & $-$ & $-$ & $-$ & $\bf 0.17_{-0.07}^{+0.13}$ & $77/91$  \\
	
	& & & & LN & $0.00104_{-0.00026}^{+0.00031}$ & $-$ & $-$ & $-$ & $-$ & $-$ & $-$ & $-$ & $\bf 0.00042_{-0.00013}^{+0.00014}$ & \\
	
	\hline

\end{tabular}%
		}
		\label{table:PDS_parameters}
		
		\begin{list}{}{}
			\item $^{\dagger}$Frozen at best fitted values as errors are not constrained.
			\item $^{\boxtimes}$Higher $\Delta \chi_{\rm dof}^{2}$ signifies very strong fundamental LFQPO (L9 component) in absence of L7 and L8 components.
		\end{list}
	\end{table}
\end{landscape}

\begin{table*}
	\caption{Best fitted QPO and harmonic characteristics obtained from different observations with {\it LAXPC20} (AS1) and {\it LAXPC10} (AS2 and AS7) of {\it AstroSat}.	All the quantities mentioned in the table have their usual meanings. See the text for details.}
	
	\renewcommand{\arraystretch}{1.4}
	\resizebox{2.1\columnwidth}{!}{%

		\begin{tabular}{l l c c c c c c c c c c c c c}
	\hline
	
	& & \multicolumn{5}{|c|}{Fundamental QPO characteristics} & & \multicolumn{4}{|c|}{Harmonic characteristics} &  \\
	
	\cline{3-7}
	\cline{9-13} \\
	
	Epoch & Energy & $\nu_{\rm QPO}$ & $\Delta \nu$ & $Q$ & $\rm Sig.$ & $rms_{\rm QPO}$ &  & $\nu_{\rm har}$ & $\Delta \nu$ & $Q$ & $\rm Sig.$ & $rms_{\rm har}$ & $rms_{\rm tot}$  & $\chi_{\rm red}^2$   \\
	
	(Date) & Range & (Hz) & (Hz) & $factor$ & ($\sigma=\frac{LN}{\rm err_{\rm neg}}^\dagger$) & ($\%$) &  & (Hz) & (Hz) & $factor$ & ($\sigma=\frac{LN}{\rm err_{\rm neg}}^\dagger$) & ($\%$) & ($\%$) & ($\chi^2/d.o.f$)\\
		
	\hline
	
	AS$1$ & $3-100$ & $1.09_{-0.02}^{+0.02}$ & $0.05_{-0.04}^{+0.03}$ & $21.80$ & $2.20$ & $9.84\pm5.84$ &  & $2.16_{-0.11}^{+0.25}$ & $0.79_{-0.27}^{+0.56}$ & $2.73$ & $3.42$ & $10.02\pm5.25$ & $22.27\pm10.39$  & $1.00$ ($44/44$) \\
	
	(02-09-2023) & $3-20$ & $1.09_{-0.02}^{+0.04}$ & $0.05_{-0.03}^{+0.02}$ & $21.80$ & $2.16$ & $11.42\pm6.27$ &  & $2.24_{-0.25}^{+0.34}$ & $0.70_{-0.32}^{+0.45}$ & $3.20$ & $3.38$ & $10.09\pm5.05$ & $19.11\pm11.94$  & $1.02$ ($45/44$) \\
	
	\hline
	
	AS$2$ & $20-100$ & $1.43_{-0.02}^{+0.02}$ & $0.20_{-0.01}^{+0.03}$ & $7.14$ & $5.46$ & $10.95\pm2.14$ &  & $2.76_{-0.04}^{+0.04}$ & $0.69_{-0.14}^{+0.16}$ & $4.00$ & $6.67$ & $6.64\pm0.31$ &  $21.88\pm1.71$  & 1.12 ($131/117$) \\
	
	(08-09-2023) & $20-40$ & $1.42_{-0.02}^{+0.02}$ & $0.22_{-0.02}^{+0.02}$ & $6.45$ & $6.51$ & $12.62\pm2.04$ &  & $2.76_{-0.04}^{+0.03}$ & $0.73_{-0.15}^{+0.18}$ & $3.78$ & $6.37$ & $7.75\pm0.63$ & $24.82\pm1.94$ & 1.10 ($128/116$) \\
	
	& $40-60$ & $1.38_{-0.01}^{+0.01}$ & $0.23_{-0.02}^{+0.02}$ & $6.00$ & $23.11$ & $10.44\pm1.28$ &  & $3.01_{-0.25}^{+0.20}$ & $1.54_{-0.74}^{+1.37}$ & $1.95$ & $2.47$ & $11.49\pm2.05$  &  $28.65\pm5.18$  & 1.06 ($135/127$) \\
	
	& $60-80$ & $1.37_{-0.01}^{+0.01}$ & $0.24_{-0.04}^{+0.04}$ & $5.71$ & $7.84$ & $5.30\pm1.22$ &  & $-$ & $-$ & $-$ & $-$ & $-$ & $26.44\pm3.06$ & 1.19 ($153/129$) \\
	
	& $80-100$ & $1.37_{-0.07}^{+0.07}$ & $0.17_{-0.07}^{+0.12}$ & $8.06$ & $3.23$ & $2.98\pm1.10$ &  & $-$ & $-$ & $-$ & $-$ & $-$ & $28.78\pm3.63$  & 0.85 ($77/91$) \\
	
	\hline
		
	AS$7$ & $20-100$ & $2.66_{-0.01}^{+0.01}$ & $0.53_{-0.02}^{+0.03}$ & $5.02$ &  $22.97$ & $9.21\pm1.35$ &  & $-$ & $-$ & $-$ & $-$ & $-$ & $17.55\pm1.69$ & 1.13 ($130/115$) \\
	
	(13-09-2023) & $20-40$ & $2.66_{-0.01}^{+0.01}$ & $0.53_{-0.02}^{+0.04}$ & $5.02$ & $23.90$ & $11.87\pm2.17$ &  & $-$ & $-$ & $-$ & $-$ & $-$ & $22.95\pm0.92$ & $1.37$ ($157/115$)$^\boxtimes$ \\
	
	& $40-60$ & $2.66_{-0.03}^{+0.04}$ & $0.63_{-0.06}^{+0.12}$ & $4.22$ & $11.26$ & $9.57\pm1.76$ &  & $-$ & $-$ & $-$ & $-$ &  $-$  & $26.46\pm2.61$  & 1.07 ($134/125$) \\
	
	& $60-80$ & $2.69_{-0.07}^{+0.06}$ & $0.69_{-0.18}^{+0.23}$ & $3.90$ & $5.41$ & $5.27\pm1.24$ &  & $-$ & $-$ & $-$ & $-$ & $-$ & $24.32\pm2.98$  & 1.19 ($159/134$) \\
		
	& $80-100$ & $2.64_{-0.08}^{+0.24}$ & $0.41_{-0.15}^{+0.33}$ & $6.44$ & $3.81$ &  $4.17\pm1.25$ &  & $-$ & $-$ & $-$ & $-$ & $-$ &  $26.62\pm3.65$ & $0.88$ ($80/91$) \\
	
	\hline
\end{tabular}%
	}
	\justify
	$^\dagger$$\rm err_{\rm neg}$ being the negative error of normalization of the fitted \texttt{Lorentzian}. $^\boxtimes$Higher $\chi_{\rm red}^2$ is observed due to the excess residuals at $\sim 2$ Hz and $\sim 35$ Hz (AS7), and we refrain using additional \texttt{Lorentzian} for modelling.
	\label{table:pds_fit}
\end{table*}

\begin{table*}
	\centering
	\caption{Best fitted and estimated spectral parameters obtained from the fitting of energy spectra with {\it NICER} and {\it AstroSat} in $0.6-40$ keV energy range. All notations have their usual meanings. The errors are computed with $90\%$ confidence level.
	}
	
	\renewcommand{\arraystretch}{1.2}
	
	\resizebox{2.0\columnwidth}{!}{%
		\begin{tabular}{l @{\hspace{0.4cm}} l @{\hspace{0.4cm}} c @{\hspace{0.3cm}} c @{\hspace{0.3cm}} c @{\hspace{0.3cm}} c c c  @{\hspace{0.3cm}} c c} \hline
			
			Model Components & Model Parameters & NI1 & AS1+NI2 & NI3 & NI4 & NI5 \\
			
			\hline
			
			& &  &  Fitted Parameters &  &  & \\ 
			
			\hline
			
			\texttt{Tbabs} & $n_{\rm H} (\times 10^{22})$ $\rm cm^{-2}$ & $0.19_{-0.01}^{+0.01}$ & $0.21_{-0.01}^{+0.01}$ & $0.20_{-0.01}^{+0.01}$ & $0.22_{-0.01}^{+0.02}$ & $0.19_{-0.01}^{+0.01}$ \\
						
			\texttt{diskbb} & $kT_{\rm in}$ (keV) & $0.32_{-0.02}^{+0.01}$ &  $0.36_{-0.01}^{+0.01}$ & $0.40_{-0.02}^{+0.02}$ & $0.60_{-0.02}^{+0.01}$ & $0.70_{-0.01}^{+0.01}$ \\
			
			& $N_{\rm diskbb}$($\times 10^{5}$) & $0.82_{-0.14}^{+0.26}$ & $1.54_{-0.13}^{+0.16}$ & $1.01_{-0.13}^{+0.18}$ & $0.13_{-0.01}^{+0.02}$ & $0.09_{-0.01}^{+0.01}$ \\ 
						
			\texttt{nthcomp} & $kT_{\rm e}$ (keV) & $3.93_{-0.42}^{+0.69}$ & $6.83_{-0.29}^{+0.30}$ & $4.26_{-0.26}^{+0.49}$ & $4^{\$}$ & $4.61^{\$}$ \\
			
			& $\Gamma_{\rm nth}$ & $1.60_{-0.02}^{+0.01}$ & $1.84_{-0.01}^{+0.01}$ & $1.80_{-0.02}^{+0.02}$ & $1.96_{-0.01}^{+0.01}$ & $2.65_{-0.02}^{+0.02}$ \\ 
						
			\texttt{gaussian} & $E$ (keV) & $-$ &  $6.41_{-0.06}^{+0.06}$ & $-$ & $6.71_{-0.01}^{+0.02}$ & $-$ \\
			
			& $\sigma$ (keV) & $-$ & $0.7^{\$}$ & $-$ & $0.30_{-0.11}^{+0.13}$  & $-$ \\
			
			& $norm$ & $-$ & $0.117_{-0.008}^{+0.008}$ & $-$ &  $0.016_{-0.006}^{+0.007}$  & $-$ \\
			
			\hline
			
			& & & Estimated Parameters & & &   \\ 
			
			\hline
			
			& $F_{\rm disc}$ ($\times 10^{-7}$) erg cm$^{-2}$ s$^{-1}$ & $0.13_{-0.01}^{+0.01}$  & $0.35_{-0.01}^{+0.01}$ & $0.36_{-0.01}^{+0.01}$ & $0.30_{-0.01}^{+0.02}$ & $0.37_{-0.01}^{+0.01}$ \\
			
			& $F_{\rm nth}$ ($\times 10^{-7}$) erg cm$^{-2}$ s$^{-1}$ & $1.27_{-0.01}^{+0.02}$  & $3.52_{-0.01}^{+0.01}$ & $1.83_{-0.02}^{+0.02}$ & $2.41_{-0.02}^{+0.02}$  &  $1.69_{-0.01}^{+0.02}$ \\
			
			& $F_{\rm bol}$ ($\times 10^{-7}$) erg cm$^{-2}$ s$^{-1}$ & $1.41_{-0.02}^{+0.02}$  & $3.98_{-0.01}^{+0.01}$ & $2.20_{-0.01}^{+0.02}$ & $2.73_{-0.02}^{+0.01}$ & $2.06_{-0.02}^{+0.03}$ \\
			
			& Disc contribution ($\%$) & $9.22$ & $8.79$ & $16.36$ & $10.99$ & $17.96$ \\
			
			& Compt. contribution ($\%$) & $90.07$ & $88.44$ & $83.18$ & $88.24$ & $82.04$ \\
			
			&  $L_{\rm bol}$ (in $L_{\rm Edd}$)  & $0.029\pm0.001$$^{*}$ & $0.082\pm0.002$$^{*}$ & $0.046\pm0.001$$^{*}$ & $0.057\pm0.003$$^{*}$  & $0.042\pm0.004$$^{*}$ \\
			
			&  & $0.32\pm0.01$$^{\dagger}$ & $0.92\pm0.01$$^{\dagger}$ & $0.51\pm0.01$$^{\dagger}$ & $0.63\pm0.02$$^{\dagger}$  &  $0.48\pm0.04$$^{\dagger}$\\
			
			
			& $\tau$ & $12.03\pm1.20$ & $6.97\pm0.22$ & $9.49\pm0.64$ & $8.66\pm1.88$  & $5.26\pm1.05$ \\
			
			& $y-par$ & $4.45\pm0.21$ & $2.60\pm0.05$ & $2.99\pm0.11$ & $2.34\pm0.15$  &  $1.01\pm0.08$ \\
			
			\hline
			
			& $\chi^2$/d.o.f ($\chi^2_{\rm red}$) & $135/136$ ($0.99$) & $223/225$ ($0.99$) & $137/135$ ($1.01$) & $124/142$ ($0.87$) & $159/161$ ($0.99$) \\
			
			\hline
			
		\end{tabular}  
	} 
	
	\begin{list}{}{}
		\item $^{\$}$Frozen parameter. $^{*}$Source distance $d=1.5$ kpc. $^{\dagger}$Source distance $d=5$ kpc. $L_{\rm Edd}=1.3\times10^{39}$ erg s$^{-1}$ for a $10 M_\odot$ black hole. 
	\end{list}
	\label{tab:spectra}
\end{table*}


\begin{thebibliography}{}
	\makeatletter
	\relax
	\def\mn@urlcharsother{\let\do\@makeother \do\$\do\&\do\#\do\^\do\_\do\%\do\~}
	\def\mn@doi{\begingroup\mn@urlcharsother \@ifnextchar [ {\mn@doi@}
		{\mn@doi@[]}}
	\def\mn@doi@[#1]#2{\def\@tempa{#1}\ifx\@tempa\@empty \href
		{http://dx.doi.org/#2} {doi:#2}\else \href {http://dx.doi.org/#2} {#1}\fi
		\endgroup}
	\def\mn@eprint#1#2{\mn@eprint@#1:#2::\@nil}
	\def\mn@eprint@arXiv#1{\href {http://arxiv.org/abs/#1} {{\tt arXiv:#1}}}
	\def\mn@eprint@dblp#1{\href {http://dblp.uni-trier.de/rec/bibtex/#1.xml}
		{dblp:#1}}
	\def\mn@eprint@#1:#2:#3:#4\@nil{\def\@tempa {#1}\def\@tempb {#2}\def\@tempc
		{#3}\ifx \@tempc \@empty \let \@tempc \@tempb \let \@tempb \@tempa \fi \ifx
		\@tempb \@empty \def\@tempb {arXiv}\fi \@ifundefined
		{mn@eprint@\@tempb}{\@tempb:\@tempc}{\expandafter \expandafter \csname
			mn@eprint@\@tempb\endcsname \expandafter{\@tempc}}}
	
	\bibitem[\protect\citeauthoryear{{Agrawal} et~al.,}{{Agrawal}
		et~al.}{2017}]{Agrawal-etal2017}
	{Agrawal} P.~C.,  et~al., 2017, \mn@doi [JAA] {10.1007/s12036-017-9451-z},
	\href {https://ui.adsabs.harvard.edu/abs/2017JApA...38...30A} {38, 30}
	
	\bibitem[\protect\citeauthoryear{{Aneesha}, {Das}, {Katoch}  \&
		{Nandi}}{{Aneesha} et~al.}{2023}]{Aneesha-etal2023}
	{Aneesha} U.,  {Das} S.,  {Katoch} T.,   {Nandi} A.,  2023, Under Review
	
	\bibitem[\protect\citeauthoryear{{Antia} et~al.,}{{Antia}
		et~al.}{2017}]{Antia-etal2017}
	{Antia} H.~M.,  et~al., 2017, \mn@doi [\apjs] {10.3847/1538-4365/aa7a0e}, \href
	{https://ui.adsabs.harvard.edu/abs/2017ApJS..231...10A} {231, 10}
	
	\bibitem[\protect\citeauthoryear{{Antia} et~al.,}{{Antia}
		et~al.}{2021}]{Antia-etal2021}
	{Antia} H.~M.,  et~al., 2021, \mn@doi [JAA] {10.1007/s12036-021-09712-8}, \href
	{https://ui.adsabs.harvard.edu/abs/2021JApA...42...32A} {42, 32}
	
	\bibitem[\protect\citeauthoryear{{Antia}, {Agrawal}, {Katoch}, {Manchanda},
		{Mukerjee}  \& {Shah}}{{Antia} et~al.}{2022}]{Antia-etal2022}
	{Antia} H.~M.,  {Agrawal} P.~C.,  {Katoch} T.,  {Manchanda} R.~K.,  {Mukerjee}
	K.,   {Shah} P.,  2022, \mn@doi [\apjs] {10.3847/1538-4365/ac6dd0}, \href
	{https://ui.adsabs.harvard.edu/abs/2022ApJS..260...40A} {260, 40}
	
	\bibitem[\protect\citeauthoryear{{Athulya}, {Radhika}, {Agrawal},
		{Ravishankar}, {Naik}, {Mandal}  \& {Nandi}}{{Athulya}
		et~al.}{2022}]{Athulya-etal2022}
	{Athulya} M.~P.,  {Radhika} D.,  {Agrawal} V.~K.,  {Ravishankar} B.~T.,  {Naik}
	S.,  {Mandal} S.,   {Nandi} A.,  2022, \mn@doi [\mnras]
	{10.1093/mnras/stab3614}, \href
	{https://ui.adsabs.harvard.edu/abs/2022MNRAS.510.3019A} {510, 3019}
	
	\bibitem[\protect\citeauthoryear{{Baby}, {Bhuvana}, {Radhika}, {Katoch},
		{Mandal}  \& {Nandi}}{{Baby} et~al.}{2021}]{Baby-etal2021}
	{Baby} B.~E.,  {Bhuvana} G.~R.,  {Radhika} D.,  {Katoch} T.,  {Mandal} S.,
	{Nandi} A.,  2021, \mn@doi [\mnras] {10.1093/mnras/stab2719}, \href
	{https://ui.adsabs.harvard.edu/abs/2021MNRAS.508.2447B} {508, 2447}
	
	\bibitem[\protect\citeauthoryear{{Belloni} \& {Altamirano}}{{Belloni} \&
		{Altamirano}}{2013}]{Belloni-etal2013}
	{Belloni} T.~M.,  {Altamirano} D.,  2013, \mn@doi [\mnras]
	{10.1093/mnras/stt500}, \href
	{https://ui.adsabs.harvard.edu/abs/2013MNRAS.432...10B} {432, 10}
	
	\bibitem[\protect\citeauthoryear{Belloni, Psaltis  \& van~der Klis}{Belloni
		et~al.}{2002}]{Belloni-etal2002}
	Belloni T.,  Psaltis D.,   van~der Klis M.,  2002, \mn@doi [The Astrophysical
	Journal] {10.1086/340290}, 572, 392
	
	\bibitem[\protect\citeauthoryear{{Belloni}, {Homan}, {Casella}, {van der Klis},
		{Nespoli}, {Lewin}, {Miller}  \& {M{\'e}ndez}}{{Belloni}
		et~al.}{2005}]{Belloni-etal2005}
	{Belloni} T.,  {Homan} J.,  {Casella} P.,  {van der Klis} M.,  {Nespoli} E.,
	{Lewin} W.~H.~G.,  {Miller} J.~M.,   {M{\'e}ndez} M.,  2005, \mn@doi [\aap]
	{10.1051/0004-6361:20042457}, \href
	{https://ui.adsabs.harvard.edu/abs/2005A&A...440..207B} {440, 207}
	
	\bibitem[\protect\citeauthoryear{{Bhuvana}, {Aneesha}, {Radhika}, {Agrawal},
		{Mandal}, {Katoch}  \& {Nandi}}{{Bhuvana} et~al.}{2023}]{Bhuvana-etal2023}
	{Bhuvana} G.~R.,  {Aneesha} U.,  {Radhika} D.,  {Agrawal} V.~K.,  {Mandal} S.,
	{Katoch} T.,   {Nandi} A.,  2023, \mn@doi [\mnras] {10.1093/mnras/stad446},
	\href {https://ui.adsabs.harvard.edu/abs/2023MNRAS.520.5828B} {520, 5828}
	
	\bibitem[\protect\citeauthoryear{{Bright}, {Farah}, {Fender}, {Siemion},
		{Pollak}  \& {DeBoer}}{{Bright} et~al.}{2023}]{Bright-etal2023}
	{Bright} J.,  {Farah} W.,  {Fender} R.,  {Siemion} A.,  {Pollak} A.,   {DeBoer}
	D.,  2023, The Astronomer's Telegram, \href
	{https://ui.adsabs.harvard.edu/abs/2023ATel16228....1B} {16228, 1}
	
	\bibitem[\protect\citeauthoryear{{Bu} et~al.,}{{Bu} et~al.}{2021}]{Bu-etal2021}
	{Bu} Q.~C.,  et~al., 2021, \mn@doi [\apj] {10.3847/1538-4357/ac11f5}, \href
	{https://ui.adsabs.harvard.edu/abs/2021ApJ...919...92B} {919, 92}
	
	\bibitem[\protect\citeauthoryear{{Chakrabarti} \& {Manickam}}{{Chakrabarti} \&
		{Manickam}}{2000}]{Chakrabarti-Manickam2000}
	{Chakrabarti} S.~K.,  {Manickam} S.~G.,  2000, \mn@doi [\apjl]
	{10.1086/312512}, \href
	{https://ui.adsabs.harvard.edu/abs/2000ApJ...531L..41C} {531, L41}
	
	\bibitem[\protect\citeauthoryear{{Chakrabarti} \& {Molteni}}{{Chakrabarti} \&
		{Molteni}}{1993}]{Chakrabarti-Molteni1993}
	{Chakrabarti} S.~K.,  {Molteni} D.,  1993, \mn@doi [\apj] {10.1086/173345},
	\href {https://ui.adsabs.harvard.edu/abs/1993ApJ...417..671C} {417, 671}
	
	\bibitem[\protect\citeauthoryear{{Chakrabarti} \& {Titarchuk}}{{Chakrabarti} \&
		{Titarchuk}}{1995}]{Chakrabarti-Titarchuk1995}
	{Chakrabarti} S.,  {Titarchuk} L.~G.,  1995, \mn@doi [\apj] {10.1086/176610},
	\href {https://ui.adsabs.harvard.edu/abs/1995ApJ...455..623C} {455, 623}
	
	\bibitem[\protect\citeauthoryear{{Chakrabarti}, {Debnath}, {Nandi}  \&
		{Pal}}{{Chakrabarti} et~al.}{2008}]{Chakrabarti-etal2008}
	{Chakrabarti} S.~K.,  {Debnath} D.,  {Nandi} A.,   {Pal} P.~S.,  2008, \mn@doi
	[\aap] {10.1051/0004-6361:200810136}, \href
	{https://ui.adsabs.harvard.edu/abs/2008A&A...489L..41C} {489, L41}
	
	\bibitem[\protect\citeauthoryear{{Das}, {Chattopadhyay}, {Nandi}  \&
		{Molteni}}{{Das} et~al.}{2014}]{Das-etal2014}
	{Das} S.,  {Chattopadhyay} I.,  {Nandi} A.,   {Molteni} D.,  2014, \mn@doi
	[\mnras] {10.1093/mnras/stu864}, \href
	{https://ui.adsabs.harvard.edu/abs/2014MNRAS.442..251D} {442, 251}
	
	\bibitem[\protect\citeauthoryear{{Dotani}, {Mitsuda}, {Makishima}  \&
		{Jones}}{{Dotani} et~al.}{1989}]{Dotani-etal1989}
	{Dotani} T.,  {Mitsuda} K.,  {Makishima} K.,   {Jones} M.~H.,  1989, \pasj,
	\href {https://ui.adsabs.harvard.edu/abs/1989PASJ...41..577D} {41, 577}
	
	\bibitem[\protect\citeauthoryear{{Draghis} et~al.,}{{Draghis}
		et~al.}{2023}]{Draghis-etal2023}
	{Draghis} P.~A.,  et~al., 2023, The Astronomer's Telegram, \href
	{https://ui.adsabs.harvard.edu/abs/2023ATel16219....1D} {16219, 1}
	
	\bibitem[\protect\citeauthoryear{{Fender}, {Belloni}  \& {Gallo}}{{Fender}
		et~al.}{2004}]{Fender-etal2004}
	{Fender} R.~P.,  {Belloni} T.~M.,   {Gallo} E.,  2004, \mn@doi [\mnras]
	{10.1111/j.1365-2966.2004.08384.x}, \href
	{https://ui.adsabs.harvard.edu/abs/2004MNRAS.355.1105F} {355, 1105}
	
	\bibitem[\protect\citeauthoryear{{Fender}, {Homan}  \& {Belloni}}{{Fender}
		et~al.}{2009}]{Fender-etal2009}
	{Fender} R.~P.,  {Homan} J.,   {Belloni} T.~M.,  2009, \mn@doi [\mnras]
	{10.1111/j.1365-2966.2009.14841.x}, \href
	{https://ui.adsabs.harvard.edu/abs/2009MNRAS.396.1370F} {396, 1370}
	
	\bibitem[\protect\citeauthoryear{{Huang} et~al.,}{{Huang}
		et~al.}{2018}]{Huang-etal2018}
	{Huang} Y.,  et~al., 2018, \mn@doi [\apj] {10.3847/1538-4357/aade4c}, \href
	{https://ui.adsabs.harvard.edu/abs/2018ApJ...866..122H} {866, 122}
	
	\bibitem[\protect\citeauthoryear{{Ingram} \& {Motta}}{{Ingram} \&
		{Motta}}{2019}]{Ingram-Motta2019}
	{Ingram} A.~R.,  {Motta} S.~E.,  2019, \mn@doi [\nar]
	{10.1016/j.newar.2020.101524}, \href
	{https://ui.adsabs.harvard.edu/abs/2019NewAR..8501524I} {85, 101524}
	
	\bibitem[\protect\citeauthoryear{{Ingram}, {van der Klis}, {Middleton}, {Done},
		{Altamirano}, {Heil}, {Uttley}  \& {Axelsson}}{{Ingram}
		et~al.}{2016}]{Ingram-etal2016}
	{Ingram} A.,  {van der Klis} M.,  {Middleton} M.,  {Done} C.,  {Altamirano} D.,
	{Heil} L.,  {Uttley} P.,   {Axelsson} M.,  2016, \mn@doi [\mnras]
	{10.1093/mnras/stw1245}, \href
	{https://ui.adsabs.harvard.edu/abs/2016MNRAS.461.1967I} {461, 1967}
	
	\bibitem[\protect\citeauthoryear{{Iyer}, {Nandi}  \& {Mandal}}{{Iyer}
		et~al.}{2015}]{Iyer-etal2015}
	{Iyer} N.,  {Nandi} A.,   {Mandal} S.,  2015, \mn@doi [\apj]
	{10.1088/0004-637X/807/1/108}, \href
	{https://ui.adsabs.harvard.edu/abs/2015ApJ...807..108I} {807, 108}
	
	\bibitem[\protect\citeauthoryear{{Katoch}, {Baby}, {Nandi}, {Agrawal}, {Antia}
		\& {Mukerjee}}{{Katoch} et~al.}{2021}]{Katoch-etal2021}
	{Katoch} T.,  {Baby} B.~E.,  {Nandi} A.,  {Agrawal} V.~K.,  {Antia} H.~M.,
	{Mukerjee} K.,  2021, \mn@doi [\mnras] {10.1093/mnras/staa3756}, \href
	{https://ui.adsabs.harvard.edu/abs/2021MNRAS.501.6123K} {501, 6123}
	
	\bibitem[\protect\citeauthoryear{{Katoch}, {Antia}, {Nandi}  \&
		{Shah}}{{Katoch} et~al.}{2023}]{Kotoch-etal2023a}
	{Katoch} T.,  {Antia} H.~M.,  {Nandi} A.,   {Shah} P.,  2023, The Astronomer's
	Telegram, \href {https://ui.adsabs.harvard.edu/abs/2023ATel16235....1K}
	{16235, 1}
	
	\bibitem[\protect\citeauthoryear{{Kennea} \& {Swift Team}}{{Kennea} \& {Swift
			Team}}{2023}]{Kennea-etal2023}
	{Kennea} J.~A.,  {Swift Team} 2023, GRB Coordinates Network, \href
	{https://ui.adsabs.harvard.edu/abs/2023GCN.34540....1K} {34540, 1}
	
	\bibitem[\protect\citeauthoryear{{Kushwaha}, {Agrawal}  \& {Nandi}}{{Kushwaha}
		et~al.}{2021}]{Kushwaha-etal2021}
	{Kushwaha} A.,  {Agrawal} V.~K.,   {Nandi} A.,  2021, \mn@doi [\mnras]
	{10.1093/mnras/stab2258}, \href
	{https://ui.adsabs.harvard.edu/abs/2021MNRAS.507.2602K} {507, 2602}
	
	\bibitem[\protect\citeauthoryear{{Li} et~al.,}{{Li} et~al.}{2023}]{Li-etal2023}
	{Li} P.~P.,  et~al., 2023, \mn@doi [\mnras] {10.1093/mnras/stad2286}, \href
	{https://ui.adsabs.harvard.edu/abs/2023MNRAS.525..595L} {525, 595}
	
	\bibitem[\protect\citeauthoryear{{Ma} et~al.,}{{Ma} et~al.}{2021}]{Ma-etal2021}
	{Ma} X.,  et~al., 2021, \mn@doi [Nature Astronomy] {10.1038/s41550-020-1192-2},
	\href {https://ui.adsabs.harvard.edu/abs/2021NatAs...5...94M} {5, 94}
	
	\bibitem[\protect\citeauthoryear{{Majumder}, {Sreehari}, {Aftab}, {Katoch},
		{Das}  \& {Nandi}}{{Majumder} et~al.}{2022}]{Majumder-etal2022}
	{Majumder} S.,  {Sreehari} H.,  {Aftab} N.,  {Katoch} T.,  {Das} S.,   {Nandi}
	A.,  2022, \mn@doi [\mnras] {10.1093/mnras/stac615}, \href
	{https://ui.adsabs.harvard.edu/abs/2022MNRAS.512.2508M} {512, 2508}
	
	\bibitem[\protect\citeauthoryear{{Majumder}, {Das}, {Agrawal}  \&
		{Nandi}}{{Majumder} et~al.}{2023}]{Majumder-etal2023}
	{Majumder} S.,  {Das} S.,  {Agrawal} V.~K.,   {Nandi} A.,  2023, \mn@doi
	[\mnras] {10.1093/mnras/stad2889}, \href
	{https://ui.adsabs.harvard.edu/abs/2023MNRAS.526.2086M} {526, 2086}
	
	\bibitem[\protect\citeauthoryear{{Makishima}, {Maejima}, {Mitsuda}, {Bradt},
		{Remillard}, {Tuohy}, {Hoshi}  \& {Nakagawa}}{{Makishima}
		et~al.}{1986}]{Makishima-etal1986}
	{Makishima} K.,  {Maejima} Y.,  {Mitsuda} K.,  {Bradt} H.~V.,  {Remillard}
	R.~A.,  {Tuohy} I.~R.,  {Hoshi} R.,   {Nakagawa} M.,  1986, \mn@doi [\apj]
	{10.1086/164534}, \href
	{https://ui.adsabs.harvard.edu/abs/1986ApJ...308..635M} {308, 635}
	
	\bibitem[\protect\citeauthoryear{{Mandal} \& {Chakrabarti}}{{Mandal} \&
		{Chakrabarti}}{2005}]{Mandal-Chakrabarti2005}
	{Mandal} S.,  {Chakrabarti} S.~K.,  2005, \mn@doi [\aap]
	{10.1051/0004-6361:20041235}, \href
	{https://ui.adsabs.harvard.edu/abs/2005A&A...434..839M} {434, 839}
	
	\bibitem[\protect\citeauthoryear{{Miller-Jones}, {Sivakoff}, {Bahramian}  \&
		{Russell}}{{Miller-Jones} et~al.}{2023}]{Miller-Jones-etal2023}
	{Miller-Jones} J.~C.~A.,  {Sivakoff} G.~R.,  {Bahramian} A.,   {Russell} T.~D.,
	2023, The Astronomer's Telegram, \href
	{https://ui.adsabs.harvard.edu/abs/2023ATel16211....1M} {16211, 1}
	
	\bibitem[\protect\citeauthoryear{{Molteni}, {Sponholz}  \&
		{Chakrabarti}}{{Molteni} et~al.}{1996}]{Molteni-etal1996}
	{Molteni} D.,  {Sponholz} H.,   {Chakrabarti} S.~K.,  1996, \mn@doi [\apj]
	{10.1086/176775}, \href
	{https://ui.adsabs.harvard.edu/abs/1996ApJ...457..805M} {457, 805}
	
	\bibitem[\protect\citeauthoryear{{Motta}, {Belloni}, {Stella},
		{Mu{\~n}oz-Darias}  \& {Fender}}{{Motta} et~al.}{2014a}]{Motta-etal2014a}
	{Motta} S.~E.,  {Belloni} T.~M.,  {Stella} L.,  {Mu{\~n}oz-Darias} T.,
	{Fender} R.,  2014a, \mn@doi [\mnras] {10.1093/mnras/stt2068}, \href
	{https://ui.adsabs.harvard.edu/abs/2014MNRAS.437.2554M} {437, 2554}
	
	\bibitem[\protect\citeauthoryear{{Motta}, {Munoz-Darias}, {Sanna}, {Fender},
		{Belloni}  \& {Stella}}{{Motta} et~al.}{2014b}]{Motta-etal2014b}
	{Motta} S.~E.,  {Munoz-Darias} T.,  {Sanna} A.,  {Fender} R.,  {Belloni} T.,
	{Stella} L.,  2014b, \mn@doi [\mnras] {10.1093/mnrasl/slt181}, \href
	{https://ui.adsabs.harvard.edu/abs/2014MNRAS.439L..65M} {439, L65}
	
	\bibitem[\protect\citeauthoryear{{Nakajima} et~al.,}{{Nakajima}
		et~al.}{2023}]{Nakajima-etal2023}
	{Nakajima} M.,  et~al., 2023, The Astronomer's Telegram, \href
	{https://ui.adsabs.harvard.edu/abs/2023ATel16206....1N} {16206, 1}
	
	\bibitem[\protect\citeauthoryear{{Nandi}, {Debnath}, {Mandal}  \&
		{Chakrabarti}}{{Nandi} et~al.}{2012}]{Nandi-etal2012}
	{Nandi} A.,  {Debnath} D.,  {Mandal} S.,   {Chakrabarti} S.~K.,  2012, \mn@doi
	[\aap] {10.1051/0004-6361/201117844}, \href
	{https://ui.adsabs.harvard.edu/abs/2012A&A...542A..56N} {542, A56}
	
	\bibitem[\protect\citeauthoryear{{Negoro} et~al.,}{{Negoro}
		et~al.}{2023}]{Negoro-etal2023}
	{Negoro} H.,  et~al., 2023, The Astronomer's Telegram, \href
	{https://ui.adsabs.harvard.edu/abs/2023ATel16205....1N} {16205, 1}
	
	\bibitem[\protect\citeauthoryear{{Nowak}}{{Nowak}}{2000}]{Nowak-etal2000}
	{Nowak} M.~A.,  2000, \mn@doi [\mnras] {10.1046/j.1365-8711.2000.03668.x},
	\href {https://ui.adsabs.harvard.edu/abs/2000MNRAS.318..361N} {318, 361}
	
	\bibitem[\protect\citeauthoryear{{Palmer} \& {Parsotan}}{{Palmer} \&
		{Parsotan}}{2023}]{Palmer-etal2023}
	{Palmer} D.~M.,  {Parsotan} T.~M.,  2023, The Astronomer's Telegram, \href
	{https://ui.adsabs.harvard.edu/abs/2023ATel16215....1P} {16215, 1}
	
	\bibitem[\protect\citeauthoryear{{Papadakis} \& {Lawrence}}{{Papadakis} \&
		{Lawrence}}{1993}]{Papadakis-etal1993}
	{Papadakis} I.~E.,  {Lawrence} A.,  1993, \mn@doi [\mnras]
	{10.1093/mnras/261.3.612}, \href
	{https://ui.adsabs.harvard.edu/abs/1993MNRAS.261..612P} {261, 612}
	
	\bibitem[\protect\citeauthoryear{{Peng} et~al.,}{{Peng}
		et~al.}{2024}]{Peng-etal2024}
	{Peng} J.-Q.,  et~al., 2024, \mn@doi [\apjl] {10.3847/2041-8213/ad17ca}, \href
	{https://ui.adsabs.harvard.edu/abs/2024ApJ...960L..17P} {960, L17}
	
	\bibitem[\protect\citeauthoryear{{Peters}, {Polisensky}, {Clarke},
		{Giacintucci}  \& {Kassim}}{{Peters} et~al.}{2023}]{Peters-etal2023}
	{Peters} W.~M.,  {Polisensky} E.,  {Clarke} T.~E.,  {Giacintucci} S.,
	{Kassim} N.~E.,  2023, The Astronomer's Telegram, \href
	{https://ui.adsabs.harvard.edu/abs/2023ATel16279....1P} {16279, 1}
	
	\bibitem[\protect\citeauthoryear{{Prabhakar}, {Mandal}, {Bhuvana}  \&
		{Nandi}}{{Prabhakar} et~al.}{2023}]{Prabhakar-etal2023}
	{Prabhakar} G.,  {Mandal} S.,  {Bhuvana} G.~R.,   {Nandi} A.,  2023, \mn@doi
	[\mnras] {10.1093/mnras/stad080}, \href
	{https://ui.adsabs.harvard.edu/abs/2023MNRAS.520.4889P} {520, 4889}
	
	\bibitem[\protect\citeauthoryear{{Radhika} \& {Nandi}}{{Radhika} \&
		{Nandi}}{2014}]{Radhika-Nandi2014}
	{Radhika} D.,  {Nandi} A.,  2014, \mn@doi [Advances in Space Research]
	{10.1016/j.asr.2014.06.039}, \href
	{https://ui.adsabs.harvard.edu/abs/2014AdSpR..54.1678R} {54, 1678}
	
	\bibitem[\protect\citeauthoryear{{Radhika}, {Nandi}, {Agrawal}  \&
		{Seetha}}{{Radhika} et~al.}{2016}]{Radhika-etal2016}
	{Radhika} D.,  {Nandi} A.,  {Agrawal} V.~K.,   {Seetha} S.,  2016, \mn@doi
	[\mnras] {10.1093/mnras/stw1239}, \href
	{https://ui.adsabs.harvard.edu/abs/2016MNRAS.460.4403R} {460, 4403}
	
	\bibitem[\protect\citeauthoryear{{Remillard} \& {McClintock}}{{Remillard} \&
		{McClintock}}{2006}]{Remillard-McClintock2006}
	{Remillard} R.~A.,  {McClintock} J.~E.,  2006, \mn@doi [\araa]
	{10.1146/annurev.astro.44.051905.092532}, \href
	{https://ui.adsabs.harvard.edu/abs/2006ARA&A..44...49R} {44, 49}
	
	\bibitem[\protect\citeauthoryear{{Ribeiro}, {M{\'e}ndez}, {de Avellar}, {Zhang}
		\& {Karpouzas}}{{Ribeiro} et~al.}{2019}]{Ribeiro-etal2019}
	{Ribeiro} E.~M.,  {M{\'e}ndez} M.,  {de Avellar} M. G.~B.,  {Zhang} G.,
	{Karpouzas} K.,  2019, \mn@doi [\mnras] {10.1093/mnras/stz2463}, \href
	{https://ui.adsabs.harvard.edu/abs/2019MNRAS.489.4980R} {489, 4980}
	
	\bibitem[\protect\citeauthoryear{{Sharma}, {Jain}  \& {Paul}}{{Sharma}
		et~al.}{2023}]{Sharma-etal2023}
	{Sharma} R.,  {Jain} C.,   {Paul} B.,  2023, \mn@doi [\mnras]
	{10.1093/mnrasl/slad110}, \href
	{https://ui.adsabs.harvard.edu/abs/2023MNRAS.526L..35S} {526, L35}
	
	\bibitem[\protect\citeauthoryear{{Sreehari}, {Ravishankar}, {Iyer}, {Agrawal},
		{Katoch}, {Mandal}  \& {Nandi}}{{Sreehari} et~al.}{2019}]{Sreehari-etal2019}
	{Sreehari} H.,  {Ravishankar} B.~T.,  {Iyer} N.,  {Agrawal} V.~K.,  {Katoch}
	T.~B.,  {Mandal} S.,   {Nandi} A.,  2019, \mn@doi [\mnras]
	{10.1093/mnras/stz1327}, \href
	{https://ui.adsabs.harvard.edu/abs/2019MNRAS.487..928S} {487, 928}
	
	\bibitem[\protect\citeauthoryear{{Sreehari}, {Nandi}, {Das}, {Agrawal},
		{Mandal}, {Ramadevi}  \& {Katoch}}{{Sreehari}
		et~al.}{2020}]{Sreehari-etal2020}
	{Sreehari} H.,  {Nandi} A.,  {Das} S.,  {Agrawal} V.~K.,  {Mandal} S.,
	{Ramadevi} M.~C.,   {Katoch} T.,  2020, \mn@doi [\mnras]
	{10.1093/mnras/staa3135}, \href
	{https://ui.adsabs.harvard.edu/abs/2020MNRAS.499.5891S} {499, 5891}
	
	\bibitem[\protect\citeauthoryear{{Tomsick} \& {Kaaret}}{{Tomsick} \&
		{Kaaret}}{2001}]{Tomsick-etal2001}
	{Tomsick} J.~A.,  {Kaaret} P.,  2001, \mn@doi [\apj] {10.1086/318683}, \href
	{https://ui.adsabs.harvard.edu/abs/2001ApJ...548..401T} {548, 401}
	
	\bibitem[\protect\citeauthoryear{{Vaughan}}{{Vaughan}}{2010}]{Vaughan2010}
	{Vaughan} S.,  2010, \mn@doi [\mnras] {10.1111/j.1365-2966.2009.15868.x}, \href
	{https://ui.adsabs.harvard.edu/abs/2010MNRAS.402..307V} {402, 307}
	
	\bibitem[\protect\citeauthoryear{{Veledina} et~al.,}{{Veledina}
		et~al.}{2023}]{Veledina-etal2023}
	{Veledina} A.,  et~al., 2023, \mn@doi [arXiv e-prints]
	{10.48550/arXiv.2309.15928}, \href
	{https://ui.adsabs.harvard.edu/abs/2023arXiv230915928V} {p. arXiv:2309.15928}
	
	\bibitem[\protect\citeauthoryear{{Wang} et~al.,}{{Wang}
		et~al.}{2021}]{Wang-etal2021}
	{Wang} Y.,  et~al., 2021, The Astronomer's Telegram, \href
	{https://ui.adsabs.harvard.edu/abs/2021ATel14613....1W} {14613, 1}
	
	\bibitem[\protect\citeauthoryear{{Zdziarski}, {Johnson}  \&
		{Magdziarz}}{{Zdziarski} et~al.}{1996}]{Zdziarski-etal1996}
	{Zdziarski} A.~A.,  {Johnson} W.~N.,   {Magdziarz} P.,  1996, \mn@doi [\mnras]
	{10.1093/mnras/283.1.193}, \href
	{https://ui.adsabs.harvard.edu/abs/1996MNRAS.283..193Z} {283, 193}
	
	\bibitem[\protect\citeauthoryear{{van der Klis}}{{van der
			Klis}}{1989}]{vanderKlis-1989}
	{van der Klis} M.,  1989, in {{\"O}gelman} H.,  {van den Heuvel} E.~P.~J.,
	eds,  NATO Advanced Study Institute (ASI) Series C Vol. 262, Timing Neutron
	Stars. p.~27, \mn@doi{10.1007/978-94-009-2273-0_3}
	
	\bibitem[\protect\citeauthoryear{{van der Klis}}{{van der
			Klis}}{1994a}]{van-der-Klis1994b}
	{van der Klis} M.,  1994a, \mn@doi [\apjs] {10.1086/192006}, \href
	{https://ui.adsabs.harvard.edu/abs/1994ApJS...92..511V} {92, 511}
	
	\bibitem[\protect\citeauthoryear{{van der Klis}}{{van der
			Klis}}{1994b}]{van-der-Klis-1994a}
	{van der Klis} M.,  1994b, \aap, \href
	{https://ui.adsabs.harvard.edu/abs/1994A&A...283..469V} {283, 469}
	
	\makeatother
\end{thebibliography}
\end{document}